\documentclass[review]{elsarticle}
\usepackage{lineno}
\modulolinenumbers[5]
\usepackage{graphicx}
\usepackage{amssymb}
\usepackage{bm,amsmath}
\usepackage{caption}
\usepackage{subeqnarray}
\usepackage{multirow}
\usepackage{lscape}
\usepackage{rotating}
\usepackage{subcaption}
\journal{Computers \& Fluids}

\begin{document}
\title{A coupled discrete unified gas-kinetic scheme for Boussinesq flows}
\author{Peng Wang\fnref{label1}}
\ead{sklccwangpeng@hust.edu.cn}
\author{Shi Tao\fnref{label1}}
\ead{lssts1013@hust.edu.cn}
\author{Zhaoli Guo\corref{cor1}\fnref{label1}}
 \ead{zlguo@hust.edu.cn}
\cortext[cor1]{Corresponding author.}

\address[label1]{State Key Laboratory of Coal Combustion, Huazhong University of Science and Technology,
Wuhan,  430074, P.R. China}
\date{\today}
\begin{abstract}
Recently, the discrete unified gas-kinetic scheme (DUGKS) [Z. L. Guo \emph{et al}., Phys. Rev. E ${\bf 88}$, 033305 (2013)] based on the Boltzmann equation is developed as a new multiscale kinetic method for isothermal flows.  In this paper, a thermal and coupled discrete unified gas-kinetic scheme is derived for the Boussinesq flows, where the velocity and temperature fields are described independently. Kinetic boundary conditions for both velocity and temperature fields are also proposed.
The proposed model is validated by simulating several canonical test cases, including the porous plate problem, the Rayleigh-b\'{e}nard convection, and the natural convection with Rayleigh number up to $10^{10}$ in a square cavity. The results show that the coupled DUGKS is of second order accuracy in space and can well describe the convection phenomena from laminar to turbulent flows. Particularly, it is found that this new scheme has better numerical stability in simulating high Rayleigh number flows compared with the previous kinetic models.

\end{abstract}

\maketitle
\section{\label{sec:level1}Introduction}

In recent years, kinetic methods have drawn particular attention as newly-developing alternative computational fluid dynamics (CFD) technology. Unlike conventional CFD methods based on direct discretizations of the Navier-Stokes equations, kinetic methods are based on the kinetic theory or the micro particle dynamics, which provides the theoretical connection between hydrodynamics and the underlying microscopic physics, and thus provides efficient tools for multiscale flows. Up to date, a variety of kinetic methods have been proposed, such as the lattice gas cellular automata (LGCA) \cite{rothman2004lattice}, the lattice Boltzmann equation (LBE) \cite{succi2001lattice,guoadvances}, the gas-kinetic scheme (GKS) \cite{xu2001gas,xu2010unified,guo2013discrete,guo2014discrete}, and the smoothed particle hydrodynamics(SPH) \cite{gingold1977smoothed}, among which the GKS and LBE are specifically designed for CFD.

Both GKS and LBE are compressible schemes for hydrodynamic equations based on gas-kinetic models, but the GKS is a finite-volume (FV) scheme originally designed for compressible flows, while LBE is a finite-difference scheme  originally designed for nearly incompressible isothermal flows with low Mach number \cite{xu2003lattice,guo2008comparative}. Later both schemes are extended to low speed thermal flows \cite{shan1997simulation, xu1999rayleigh, guo2002coupled,shi2006finite,guo2007thermal,he1998novel}. Generally, thermal effects in nearly incompressible flows can lead to large compressibility errors for a compressible scheme \cite{xu1999rayleigh}, and in order to reduce such difficulty, the mass and momentum equations are decoupled from the energy equation. Such strategy has been adopted in both GKS and LBE methods  \cite{shan1997simulation, xu1999rayleigh,guo2002coupled,shi2006finite, guo2007thermal,he1998novel,wang2013lattice,kao2007simulating,shi2004thermal,watanabe2004flow,shu2005lattice}.

Recently, starting from the Boltzmann equation, a discrete unified gas-kinetic scheme (DUGKS) was proposed for isothermal flows in all Knudsen regimes \cite{guo2013discrete}. The DUGKS is a FV method, which combines the advantages of the GKS in its flux modeling and the LBE methods in its expanded Maxwellian distribution function and discrete conservative collision operator. In addition, the DUGKS has the asymptotic preserving (AP) property in capturing both rarefied and Navier-Stokes equations solutions in the corresponding flow regime \cite{xu2010unified}.

Particularly, although sharing a common kinetic origin, some distinctive features also exist between DUGKS and LBE methods. First, the standard LBE methods can be viewed as special finite-difference schemes, while the DUGKS is a FV one. Second, although both LBE and DUGKS methods evolve in discrete phase space (physical and particle velocity space) and discrete time, in the LBE methods the phase space and time step are coupled due to the particle motion from one node to another one within a time step, whereas the DUGKS does not suffer from this restriction and the time step is fully determined by the Courant-Friedrichs-Lewy (CFL) condition. Third, the streaming process coupling between the discrete velocity and the underlying regular lattice in LBE makes it quite difficult to be extended to non-uniform mesh, while for the DUGKS the non-uniform mesh can be easily employed without additional efforts. More importantly, there are modeling differences in LBE and DUGKS in the particle evolution process. The LBE separates the particle streaming and collision process in its algorithm development. But, the particle transport and collision are fully coupled in DUGKS. This dynamic difference determines the solution deviation in their flow simulations. Consequently, it has been demonstrated that the DUGKS can achieve  identical accurate results for the incompressible flows in comparison with the LBE methods, but is more robust and stable \cite{wangcomparative}.

Although the DUGKS has such distinctive features, the original DUGKS is only designed for isothermal flows which limits its applications \cite{guo2013discrete}. The motivation of  this work is to develop a DUGKS for near incompressible thermal flows under the decoupling strategy, where the velocity and temperature fields are described by two respective DUGKS models which are coupled under the Boussinesq assumption. Kinetic boundary conditions are also proposed for both the velocity and temperature fields. To validate the performance of the coupled DUGKS, two-dimensional (2D) porous plate problem, the Rayleigh-b\'{e}nard problem and the natural convection in a square cavity at Rayleigh number from $10^3$ up to $10^{10}$ are simulated.

 The rest of this paper is organized as follows. In Sec.~\ref{kinetic model}, the coupled DUGKS and the kinetic boundary conditions for velocity and temperature fields are developed, some numerical tests are made in Sec.~\ref{numerical} to validate the performance of the new scheme, and  a brief summary is presented in Sec.~\ref{conclusion}.

 \section{COUPLED DISCRETE UNIFIED GAS-KINETIC SCHEME}
\label{kinetic model}
In this section, we first introduce the gas-kinetic model for the Boussinesq flows. Then, the DUGKS based on the model will be derived for velocity and  temperature fields, respectively.  The two evolution equations are coupled based on the Boussinesq assumption. The kinetic boundary conditions and algorithm for velocity and temperature fields are introduced finally.

\subsection{Gas-kinetic model for Boussinesq flows}
In this subsection, we are going to introduce the gas-kinetic model for the following incompressible Navier-Stokes equations with the thermal effects \cite{xu1999rayleigh}:
\begin{equation}\begin{split}\label{NS}
\frac{\partial \rho}{\partial t}+\nabla\cdot(\rho\bm{u})&=0,\\
\frac{\partial \bm u}{\partial t}+\bm{u}\cdot\nabla\bm{u}&=-\frac{\nabla p}{\rho}+\nu\nabla^2\bm{u}-\bm a,\\
\frac{\partial T}{\partial t}+\nabla\cdot \left(T\bm{u}\right)&=\nabla\cdot\left(\kappa\nabla{T}\right),
\end{split}\end{equation}
where $\rho$, $\bm u$, $T$, $p$ are the density, velocity, temperature and pressure of the flow fluid, respectively, $\kappa$ is the coefficient of thermal conductivity, and $\bm{a}$ is the accelerated velocity of the external force. With the Boussinesq approximation, the force term can be written as:
\begin{equation}\label{BOSS}
\rho \bm a=\rho g_0 \beta\left(T-T_0\right)\bm y,
\end{equation}
where $g_0$ is the gravitational constant, $T_0$ is the reference temperature, $\bm y$ is the unit vector in the vertical direction, and $\beta$ is the coefficient of volume expansion. The gas-kinetic equations corresponding to the above equations can be constructed as \cite{xu1999rayleigh}:
\begin{equation}\label{BGK1}
 \frac{\partial f}{\partial t}+{\bm \xi}\cdot\nabla_x f+{\bm a}\cdot\nabla_\xi f=\Omega\equiv\frac{f^{eq}-f}{\tau_v},
\end{equation}
\begin{equation}\label{BGK2}
 \frac{\partial g}{\partial t}+{\bm \xi}\cdot\nabla g=\Psi\equiv\frac{g^{eq}-g}{\tau_c},
\end{equation}
where $f$ and $g$ are the gas distribution functions for the velocity and temperature fields, respectively, and  $f^{eq}$ and $g^{eq}$ are the corresponding  equilibrium states. Both $f$ and $g$ are functions of space $\bm x$, time $t$, and particle velocity ${\bm \xi}$, and the particle collision time $\tau_v$ and $\tau_c$ are related to the viscosity and the heat conduction coefficients, respectively.  The equilibrium states  $f^{eq}$ and $g^{eq}$ take the following forms:
\begin{equation}\label{MAX1}
f^{eq}=\frac{\rho}{{(2\pi R T_1)}^{D/2}}\exp\left(-\frac{({\bm \xi}-{\bm u})^{2}}{2RT_1}\right),
\end{equation}
\begin{equation}\label{MAX2}
g^{eq}=\frac{T}{{(2\pi R T_2)}^{D/2}}\exp\left(-\frac{({\bm \xi}-{\bm u})^{2}}{2RT_2}\right),
\end{equation}
where $R$ is the gas constant,  $T_1$ and $T_2$ are the constant variances which determine the artificial sound speed of the velocity. For continuum flows, the external force term can be approximated as \cite{shan1997simulation}:
\begin{equation}\label{force}
{\bm a}\cdot\nabla_\xi f\approx{\bm a}\cdot\nabla_\xi f^{eq}=-\frac{\bm{a}\cdot\left(\bm \xi-\bm u\right)}{RT_1}f^{eq}.
\end{equation}

Using the Chapman-Enskog expansion, the hydrodynamic equations (Eq.~\eqref{NS}) can be obtained from Eq.~\eqref{BGK1} and Eq.~\eqref{BGK2} exactly in the incompressible limit, with the viscosity coefficient
\begin{equation}\label{nu}
\nu=\tau_v R T_1,
\end{equation}
and the heat conduction coefficient
\begin{equation}\label{kappa}
\kappa=\tau_c R T_2.
\end{equation}
Therefore, the Prandtl number $\text{Pr}$ can be modified by choosing appropriate $\tau_v$, $\tau_c$, $T_1$, and $T_2$, which gives:
  \begin{equation}\label{force}
\text{Pr}=\frac{\nu}{\kappa}=\frac{\tau_v}{\tau_c}\frac{T_1}{T_2}.
\end{equation}
\subsection{DUGKS for velocity}
\label{velocity}
Unlike most of other kinetic methods, the DUGKS is a semi-Lagrangian FV scheme, where the
evolution process is under the Eulerian framework and the flux construction at the cell interface is based on the
Lagrangian perspective. It is noted that the original DUGKS does not consider external force term. Hence, in order to model the velocity with a body force, we rewrite Eq.~\eqref{BGK1} as:
\begin{equation}\label{BGK3}
 \frac{\partial f}{\partial t}+{\bm \xi}\cdot\nabla_x f=\bar{\Omega}\equiv \Omega+ F,
\end{equation}
where
\begin{equation}\label{force1}
F=a\cdot\nabla_{\xi}f\approx\frac{\bm{a}\cdot\left(\bm \xi-\bm u\right)}{RT_1}f^{eq}.
\end{equation}
Notice that $\int F d\bm \xi=0$, and $\int F\bm \xi d\bm \xi=\rho \bm a$.

 As a FV scheme, in the DUGKS the computational domain is divided into a set of control volumes.  Integrating Eq.~\eqref{BGK3} over a control volume $V_j$ centered at $\bm x_j$ from time $t_{n}$ to $t_{n+1}$ (the time step $\Delta t=t_{n+1}-t_n$ is assumed to be a constant in the present work), and using the midpoint rule for the integration of the flux term at the cell interface and trapezoidal rule for the collision term inside each cell , the evolution equation for velocity can be written as \cite{guo2013discrete}:
\begin{equation}\label{update}
\tilde{f}^{n+1}_j=\tilde{f}^{+,n}_j-\frac{\Delta t}{|V_j|}F^{n+1/2},
\end{equation}
where
\begin{equation}\label{microflux}
 F^{n+1/2}=\int_{\partial V_j}\left(\bm \xi \cdot \bm n\right)f\left(\bm x,t_{n+1/2}\right)d{\bm S}
\end{equation}
is the microflux across the cell interface, ${\bm n}$ is the unit vector normal to the cell interface and
\begin{equation}\label{aux}
\tilde{f}=f-\frac{\Delta t}{2}\bar{\Omega},\hspace{2mm} \tilde{f}^+=f+\frac{\Delta t}{2}\bar{\Omega}
\end{equation}
are the auxiliary distributions related to the original distribution function $f$ and the equilibrium distribution function $f^{eq}$.  Clearly, the evolution process is Eulerian. Based on the compatibility condition and the relation of $f$ and $\tilde{f}$, the density  and velocity can be computed by:
\begin{equation}\label{rho}
\rho=\int\tilde{f}d\bm\xi,\hspace{5mm} \rho \bm u=\int\bm{\xi}\tilde{f}d\bm\xi+\frac{\Delta t}{2}\rho \bm{a}.
\end{equation}

The key procedure in updating $\tilde{f}$ is to evaluate the microflux $F^{n+1/2}$, which can be solely determined by the gas distribution function $f(\bm x,t_{n+1/2})$ at the cell interface. The Lagrangian perspective is applied in the construction of $f(\bm x,t_{n+1/2})$. To this end, Eq.~\eqref{BGK3} is integrated within a half time step $h=\Delta t /2$
along the characteristic line with the end point $(\bm{x}_b)$ located at the cell interface, and the trapezoidal rule is applied to evaluate the collision term,
\begin{equation}\label{BGKaa}
f({\bm x}_b,{\bm \xi},t_n+h)-f({\bm x}_b-\bm{\xi}h,{\bm \xi},t_n)=\frac{h}{2}\left[\bar{\Omega}(\bm{x}_b,\bm{\xi},t_n+h)+\bar{\Omega}({\bm x}_b-\bm{\xi}h,{\bm \xi},t_n) \right].
\end{equation}
In order to remove the implicity of Eq.~\eqref{BGKaa}, two auxiliary distribution functions are introduced,
  \begin{equation}\label{aux2}
   \bar{f}=f-\frac{h}{2}\bar{\Omega}, \hspace{2mm}{\bar f}^+=f+\frac{h}{2}\bar{\Omega}.
\end{equation}
Note that the particle collision effect is included in the above evaluation of the interface
gas distribution function. This is the key for the success of the DUGKS \cite{wangcomparative}. With the newly introduced distribution functions, Eq.~\eqref{BGKaa} can be rewritten as:
 \begin{equation}\label{BGK_face3a}
{\bar{f}}({\bm x}_b,{\bm \xi},t_n+h)={\bar{f}}^+\left({\bm x}_b-\bm{\xi}h,{\bm \xi},t_n\right).
\end{equation}
For smooth flows, ${\bar{f}}^+\left({\bm x}_b-\bm{\xi}h,{\bm \xi},t_n\right)$ can be approximated by its Taylor expansion around the cell interface $\bm{x}_b$,
\begin{equation}\label{BGK_face3b}
{\bar{f}}^+\left({\bm x}_b-\bm{\xi}h,{\bm \xi},t_n\right)={\bar{f}}^+\left({\bm x}_b,{\bm \xi},t_n\right)-h{\bm \xi}\cdot{\bm \sigma}_b,
\end{equation}
where ${{\bm \sigma}}_b=\nabla {\bar{f}}^+({\bm x}_b,{\bm \xi},t_n) $. Based on Eqs.~\eqref{BGK_face3a} and \eqref{BGK_face3b}, one can get:
 \begin{equation}\label{BGK_face3}
{\bar{f}}({\bm x}_b,{\bm \xi},t_n+h)={\bar{f}}^+\left({\bm x}_b,{\bm \xi},t_n\right)-h{\bm \xi}\cdot{\bm \sigma}_b.
\end{equation}
Then, based on the compatibility condition and the relation of $f$ and $\bar{f}$, the density and velocity at the cell interface can be obtained:
\begin{equation}\label{rho1}
\rho=\int\bar{f}d\bm\xi,\hspace{5mm} \rho \bm u=\int\bm{\xi}\bar{f}d\bm\xi+\frac{h}{2}\rho \bm{a},
\end{equation}
from which the equilibrium distribution function $f^{eq}\left({\bm x}_b,{\bm \xi},t_n+h\right)$ at the cell interface can be obtained. Therefore, based on Eq.~\eqref{aux2} and the obtained equilibrium state, the original distribution function at the cell interface can be extracted from $\bar{f}$,
\begin{equation}\label{original}
f(\bm{x}_b, t_n+h)=\frac{2\tau_v}{2\tau_v+h}\bar{f}\left(\bm{x}_b,t_n+h\right)+\frac{h}{2\tau_v+h}f^{eq}\left(\bm{x}_b,t_n+h\right)+\frac{\tau_v h}{2\tau_v+h}F,
\end{equation}
from which the interface flux can be evaluated.

As a result, the update of the distribution function $\tilde{f}$ can be done according to Eq.~\eqref{update}.  In practical computations, we only need to track the evolution of $\tilde{f}$, and the required variables in the evolution are:
\begin{equation}\label{relationv1}
\bar{f}^+=\frac{2\tau_v-h}{2\tau_v+\Delta t}\tilde{f}+\frac{3h}{2\tau_v+\Delta t}f^{eq}+\frac{3h\tau_v}{2\tau_v+\Delta t}F,
\end{equation}
\begin{equation}\label{relationv3}
\tilde{f}^+=\frac{4}{3}\bar{f}^+-\frac{1}{3}\tilde{f}.
\end{equation}

Up to this point, the scheme is designed with the continuous velocity space $\bm \xi$, while the discrete particle velocities ${\bm \xi}_i$ is  employed in DUGKS. Similar to LBE, for low Mach number flows, the Maxwellian distribution can be approximated by its Taylor expansion around zero particle velocity,
\begin{equation}\label{equilibrium}
{f^{eq}_i}=W_i\rho\left[1+\frac{\bm{\xi}\cdot\bm{u}}{RT_1}+ \frac{(\bm{\xi}\cdot\bm{u})^2}{2(RT_1)^2}-\frac{\mid\bm{u}\mid^2}{2RT_1} \right],
\end{equation}
where $f^{eq}_i=w_if^{eq}(\xi_i)$, $w_i=W_i(2\pi RT_1)^{D/2}\text{exp}\left( \frac{\mid \bm{\xi_i}\mid^2}{2RT_1}\right)$, and $W_i$ is the weight coefficients corresponding to the abscissas $\bm{\xi_i}$.

\subsection{DUGKS for temperature field}
\label{temperature}
A DUGKS model for Eq.~\eqref{BGK2} can be constructed similarly. The Eq.~\eqref{BGK2} is first integrated  at the same control volume $V_j$ from $t_n$ to $t_{n+1}$, and then the same integration rules are employed to approximate the convection term and collision term as that in the velocity, one can get:
\begin{equation}\label{BGKT}
 g_j^{n+1}- g_j^{n}+\frac{\Delta t}{|V_j|}\digamma^{n+1/2}=\frac{\Delta t}{2}\left[{\Psi}_j^{n+1}+{\Psi}_j^{n}\right],
\end{equation}
where  $\digamma$ is the microflux,
\begin{equation}\label{microfluxg}
 \digamma^{n+1/2}=\int_{\partial V_j}\left(\bm \xi \cdot \bm n \right)g\left(\bm x,t_{n+1/2}\right)d{\bm S},
\end{equation}
${g}^{n}_j$ and ${\Psi}_j^{n}$  are the cell-averaged values of the distribution function and the collision term, respectively, e.g.,
\begin{equation}\label{average_u}
{g}^{n}_j=\frac{1}{|V_j|}\int_{V_j}g\left(\bm x,t_{n}\right)d{\bm x}.
\end{equation}
Two auxiliary distribution functions are introduced to remove the implicity in Eq.~\eqref{BGKT}, then the evolution equation of  Eq.~\eqref{BGK2} can be written as:
\begin{equation}\label{updateT}
\tilde{g}^{n+1}_j=\tilde{g}^{+,n}_j-\frac{\Delta t}{|V_j|}\digamma^{n+1/2},
\end{equation}
where
\begin{equation}\label{aux}
\tilde{g}=g-\frac{\Delta t}{2}{\Psi},\hspace{2mm} \tilde{g}^+=g+\frac{\Delta t}{2}{\Psi},
\end{equation}
from which the temperature can be computed as:
\begin{equation}\label{tem}
T=\int\tilde{g}d\bm\xi.
\end{equation}

In order to evaluate the microflux $\digamma$, we also integrate Eq.~\eqref{BGK2} within a half time step $h$ along the characteristic line with the end point $\bm{x}_b$ at the cell interface, and use the trapezoidal rule to evaluate the collision term,
 \begin{equation}\label{BGK_faceT}
g\left({\bm x}_b,{\bm \xi},t_n+h\right)-g\left({\bm x}_b-{\bm \xi}h,{\bm \xi},t_n\right)=\frac{h}{2}\left[{\Psi}({\bm x}_b,t_n+h)+{\Psi}({\bm x}_b-{\bm \xi}h,t_n)\right].
\end{equation}
Also another two new distribution functions are introduced to remove the implicity in the above equation,
  \begin{equation}\label{auxT2}
 \bar{g}=g-\frac{h}{2}{\Psi}, \hspace{2mm}{\bar g}^+=g-\frac{h}{2}{\Psi}.
\end{equation}
Then, Eq.~\eqref{BGK_faceT}  can be rewritten as:
\begin{equation}\begin{split}\label{BGK_face1T}
\bar{g}\left({\bm x}_b,{\bm \xi},t_n+h\right)&={\bar{g}}^+\left({\bm x}_b-{\bm \xi}h,{\bm \xi},t_n\right)\\
&={\bar{g}}^+({\bm x}_b,{\bm \xi},t_n)-h{\bm \xi}\cdot{ \bar{\bm\sigma}}_b,
\end{split}\end{equation}
where ${\bar{\bm \sigma}}_b=\nabla {\bar{g}}^+\left({\bm x}_b,{\bm \xi},t_n\right) $ and the Taylor expansion is made around the cell interface $\bm x_b$.  Based on Eq.~\eqref{auxT2} and the compatibility condition,  the temperature at the cell interface can be computed as:
\begin{equation}\label{tem}
T=\int\bar{g}d\bm\xi.
\end{equation}
Together with the conserved variables in velocity, the equilibrium distribution function $g^{eq}({\bm x}_b,{\bm \xi},t^n+h)$ can be fully determined. Then, the original distribution function can be obtained,
\begin{equation}\label{originalT}
g(\bm{x}_b, t_n+h)=\frac{2\tau_c}{2\tau_c+h}\bar{g}\left(\bm{x}_b,t_n+h\right)+\frac{h}{2\tau_c+h}g^{eq}\left(\bm{x}_b,t_n+h\right),
\end{equation}
from which the interface numerical flux $\digamma$  can be evaluated.

In computation, it only needs to follow the evolution of $\tilde{g}$ in Eq.~\eqref{updateT}.  The required variables for its evolution are determined by
\begin{equation}\label{relationt4}
\bar{g}^+=\frac{2\tau_c-h}{2\tau_c+\Delta t}\tilde{g}+\frac{3h}{2\tau_c+\Delta t}g^{eq},
\end{equation}
\begin{equation}\label{relationt5}
\tilde{g}^+=\frac{4}{3}\bar{g}^+-\frac{1}{3}\tilde{g}.
\end{equation}

 In the present work, the DUGKS for temperature field  uses the same discrete velocity set as that for velocity field, and the expanded discrete equilibrium distribution function can be written as:
\begin{equation}\label{equilibriumg}
{g^{eq}_i}=W_i T\left[1+\frac{\bm{\xi}\cdot\bm{u}}{RT_2}+ \frac{(\bm{\xi}\cdot\bm{u})^2}{2(RT_2)^2}-\frac{\mid\bm{u}\mid^2}{2RT_2} \right],
\end{equation}
where $g^{eq}_i=w_ig^{eq}(\xi_i)$, $w_i=W_i(2\pi RT_2)^{D/2}\text{exp}\left( \frac{\mid \bm{\xi_i}\mid^2}{2RT_2}\right)$, and $W_i$ is the weight coefficient corresponding to the abscissas $\bm{\xi_i}$.

\subsection{Kinetic boundary conditions}
\label{boundary}
Boundary condition plays an important role in kinetic models in that they will influence their accuracy and stability \cite{mei1999accurate,zou1997pressure}.
In the previous study, two types of kinetic boundary conditions for velocity without external force have been specified in DUGKS \cite{guo2013discrete}, among which the bounce-back rule that assumes a particle just reverses its velocity after hitting the wall is presented for no-slip boundary,
\begin{equation}\label{bounce_flow}
f(\bm x_w,\bm\xi_i,t+h)=f(\bm x_w,-\bm\xi_i,t+h)+2\rho_w\frac{W_i}{w_i}\frac{\bm{\xi}_i\cdot\bm{u_w}}{RT_1},\hspace{5mm}\bm \xi_i\cdot \bm n>0,
\end{equation}
where  $\rho_w$ and $\bm{u}_w$ are the density and velocity at the wall, respectively, and $\bm n$ is the unit vector normal to the wall pointing to the cell. However, in practical calculations, the walls are fixed at the cell interface, thereby $\bar{f}$ and $\bar{g}$ are needed when dealing with the boundary conditions.  For the velocity field, we should give $\bar{f}(\bm x_w,\bm\xi_i,t)$ other than $f(\bm x_w,\bm\xi_i,t)$ in the boundary condition, and Eq.~\eqref{bounce_flow} can be rewritten as:
 \begin{equation}\label{bounce_flow_F}
\bar{f}(\bm x_w,\bm\xi_i,t+h)=\bar{f}(\bm x_w,-\bm\xi_i,t+h)+2\rho_w\frac{W_i}{w_i}\frac{\bm{\xi}_i\cdot\bm{u}_w'}{RT_1},\hspace{5mm}\bm \xi_i\cdot \bm n>0,
\end{equation}
where $\bm{u}_w'=\bm{u}_w-(h/2)\bm{a}$. For nearly incompressible flow, $\rho_w$ can be approximated well by the constant average density.

As for temperature field, we consider two types of boundaries i.e., wall with a fixed temperature and adiabatic wall. First, for the constant temperature boundary, the distribution function $\bar{g}(\bm x_w,\bm\xi_i,t)$ for particle leaving the wall can be constructed as \cite{zhang2012general}:

\begin{equation}\label{bounce_Tc}
\bar{g}(\bm x_w,\bm\xi_i,t+h)=-\bar{g}(\bm x_w,-\bm\xi_i,t+h)+2\frac{W_i}{w_i}T_w \left[1.0+\frac{(\bm{\xi}\cdot\bm{u}_w)^2}{2(RT_2)^2}-\frac{\mid\bm{u}_w\mid^2}{2RT_2} \right],\bm \xi_i\cdot \bm n>0,
\end{equation}
where  $T_w$  is the wall temperature.  Second, the adiabatic boundary, which is a Neumann boundary condition, can be realized by the bounce-back rule \cite{wang2013lattice},
 \begin{equation}
 \label{bounce_Ta}
 \bar{g}(\bm x_w,\bm\xi_i,t+h)=\bar{g}(\bm x_w,-\bm\xi_i,t+h).
 \end{equation}
The distribution function  $\bar{g}(\bm x_w,\bm\xi_i,t+h)$ for particle moving towards the wall, i.e., $\bm \xi_i\cdot \bm n\leq0$ , can be constructed following the procedure described in Sec.~\ref{temperature}.

\subsection{Algorithm}
In this subsection, we list the computational procedures for the updating of the discrete distribution functions in both the velocity and temperature fields. In the computation, the weight coefficients $w_i$ are absorbed into the discrete functions, i.e.,
\begin{equation}
f_i=w_if(\xi_i),\hspace{1cm} g_i=w_ig(\xi_i).
\end{equation}
Note that the distributions $\tilde{f}$ and $\tilde{g}$ are recorded instead of the original one, respectively, so that the macroscopic variables can be evaluated as:
\begin{equation}\label{drho}
\rho=\sum\limits_{i}\tilde{f}_i,\hspace{5mm} \rho \bm u=\sum\limits_{i}\bm{\xi}_i\tilde{f}_i+\frac{\Delta t}{2}\rho\bm{a}, \hspace{5mm} T=\sum\limits_{i}\tilde{g}_i.
\end{equation}

The updates of $\tilde{f}$ and $\tilde{g}$ are same as that for the continuous cases presented in Secs.~\ref{velocity} and \ref{temperature}. Specially with initialized $\tilde{f}_{j,i}^0$ and $\tilde{g}_{j,i}^0$ in all cells centered at $\bm{x}_j$, the procedure of the DUGKS at each time $t_n$ reads as follows:

(1) Compute the distribution functions $\bar{f}_{j,i}^{+,n}$ ( Eq.~\eqref{relationv1}) and $\bar{g}_{j,i}^{+,n}$ (Eq.~\eqref{relationt4}) in each cell.

(2) Compute the distribution functions $\bar{f}_{i}^{n+1/2}(\bm{x}_b)$ (Eqs.~\eqref{BGK_face3}) and  $\bar{g}_{i}^{n+1/2}(\bm{x}_b)$ (Eq.~\eqref{BGK_face1T}).

(3) Compute the original distribution functions  $f_{i}^{n+1/2}(\bm{x}_b)$ (Eq.~\eqref{original}) and $f_{i}^{n+1/2}(\bm{x}_b)$ (Eq.~\eqref{originalT}).

(4) Compute the microflux across the cell interfaces from $f_{i}^{n+1/2}(\bm{x}_b)$ (Eq.~\eqref{microflux}) and $f_{i}^{n+1/2}(\bm{x}_b)$ (Eq.~\eqref{microfluxg}).

(4) Update the distribution functions $\bar{f}_{j,i}^{+,n}$ and $\bar{g}_{j,i}^{+,n}$ via Eqs.~\eqref{update} and \eqref{updateT}, respectively, where $\tilde{f}^{+}$ and $\tilde{g}^{+}$ are computed respectively according to Eqs.~\eqref{relationv3} and \eqref{relationt5}.
\section{NUMERICAL RESULTS}
\label{numerical}
  In this section, several numerical simulations are conducted to validate the proposed model, including the porous plate problem, the Rayleigh-b\'{e}nard convection, and the natural convection in a square cavity. In our simulations,  $T_1=T_2=T_0$ are taken for $f^{eq}$ and $g^{eq}$ although they can be different in theory,  and the three-point Gauss-Hermite quadrature is used to evaluate the moments, which yields the following discrete velocities and associated weights,
 \begin{equation}\begin{split}
 &\bm{\xi_{-1}}=-\sqrt{3RT_0}, \hspace{2mm}  \bm{\xi_{0}}=0, \hspace{2mm}  \bm{\xi_{1}}=\sqrt{3RT_0},\\
 &W_0=2/3, \hspace{2mm} W_{\pm 1}=1/6.
 \end{split}\end{equation}
 For the two-dimensional problems considered in this paper, the discrete velocities and weights used in the DUGKS are generated using the tensor product method  \cite{guo2013discrete}.

In the following simulations, the Mach number is defined as $Ma=U/c_s$, where $U=\sqrt{g_0\beta\Delta T H}$ is the characteristic velocity of the flow, $c_s=\sqrt{RT_0}$ is the speed of sound, $\Delta T$ is the temperature difference, and $H$ is the characteristic length;  the collision time are determined by $\tau_v=\mu /p$ and $\tau_c=\tau_v/\text{Pr}$, where $\mu$ is the dynamic viscosity,  $p=\rho RT_0$ is the pressure; the time step $\Delta t$ is determined by the CFL number, i.e., $\Delta t=\eta\Delta x_{min}/C$, where $\eta$ is the CFL number, $\Delta x_{min}$ is minimum grid spacing, and $C$ is on the order of the maximal discrete velocity; uniform meshes are adopted for most of the test cases except for the turbulent natural convection flow where a non-uniform mesh is applied with the requirement of the local accuracy; we set $g_0\beta=0.1$ and $\eta=0.5$ in what follows unless otherwise stated; for the steady state cases, the working criterion is defined by:
\begin{equation}\label{work_state}
\begin{split}
&\frac{\sqrt{\sum \left\|\bm{u}(t)-\bm{u}(t-1000\Delta t)\right\|^2}}{\sqrt{\sum  \|\bm{u}(t)\|^2}}<10^{-12}\\
& {max} \hspace{2mm}| T(t)-T(t-1000\Delta t) | < 10^{-6}.
\end{split}
\end{equation}
Note that all the parameters employed in our simulations are non-dimensional.

\subsection{Porous plate problem}
We first study the accuracy of the coupled DUGKS model by simulating the porous plate problem, which has an analytic solution.
The problem considered is a channel flow where the upper cool plate with a constant temperature $T_c$ moves with a constant velocity $u_0$, and a constant normal flow  is injected with a constant velocity $v_0$ through the bottom hot plate with a constant temperature $T_h$ and withdraw with the same rate from the upper plate. This problem models a fluid being sheared
between two plates through which an identical fluid is being injected normal to the shearing direction.
The analytic solutions of horizontal velocity and temperature for this problem in steady state are given by \cite{guo2002coupled}:
\begin{equation}\label{velocity_porous_u}
u=u_0\left(\frac{e^{\text{Re}\hspace{1mm}y/H}-1}{e^{\text{Re}}-1}\right),
\end{equation}
\begin{equation}\label{velocity_porous_T}
T=T_h-\Delta T\left(\frac{e^{\text{Pr}\hspace{1mm} \text{Re}\hspace{1mm} y/H}-1}{e^{\text{Pr} \hspace{1mm}\text{Re}}-1}\right),
\end{equation}
where $\text{Re}$ is the Reynolds number based on the inject velocity $v_0$, $\Delta T=T_h-T_c$ is the temperature difference between the hot and cool walls, and $\text{Ra}=g_0\beta\Delta T H^3/(\nu\kappa)$ is the Rayleigh number.
\begin{figure}[!htb]
\centering
\includegraphics[width=0.5\textwidth]{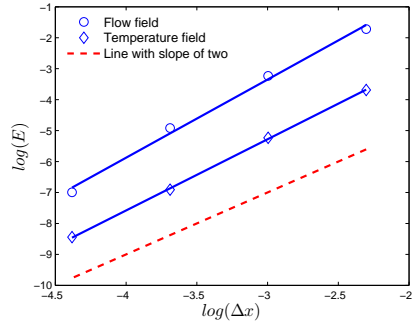}
\caption{Numerical error versus grid size for the porous plate flow. The slope of the dashed line is 2.}\label{error}
\end{figure}
 \begin{figure}[!htb]
 \centering
\includegraphics[width=0.5\textwidth]{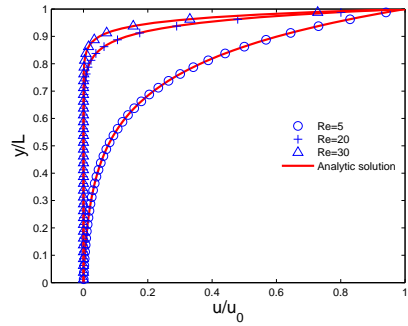}
\caption{Velocity profiles at  $\text{Re}=5, 20 $ and $30$ of the porous plate problem with $\text{Pr}=0.71, \text{Ra}=100$.}\label{multiv}
\end{figure}

In our simulations, we set $\Delta T=1.0$ and $u_0=0.1$; the length and the height of the channel are $L=2$ and $H=1$, respectively.
Boundary conditions presented in Sec.~\ref{boundary} are applied to the plates,  and periodic boundary conditions are applied to inlet and outlet of the channel. In order to evaluate the accuracy of the coupled DUGKS, a set of simulations with different mesh resolutions are conducted.  We  use $\text{Pr}=0.71$, $\text{Re}=10$, $\text{Ra}=100$, and the grid size varies from $1/10$ to $1/80$, $RT_0=100$ so that the time step $\Delta t$ is small enough to reduce the time error in the evaluation of spatial accuracy and the flow can be treated as incompressible. Moreover, the CFL number is adjusted to keep the time step constant. The relative global error in velocity and temperature fields is defined as:
 \begin{equation}
 \label{err}
 E=\frac{\sqrt{\sum\left\|{A}(\bm{x})-{A}^*(\bm{x})\right\|^2}}{\sqrt{\sum \|{A}^*(\bm{x})\|^2}},
 \end{equation}
 where $A$  ($u$ or $T$) is the numerical result, $A^*$ is the analytic solution given by Eqs.~\eqref{velocity_porous_u} and \eqref{velocity_porous_T}. The errors in velocity and temperature fields are shown in Fig.~\ref{error}, which shows that the coupled DUGKS is of second-order accuracy in space.

\begin{figure}[!htb]
\centering
\includegraphics[width=0.5\textwidth]{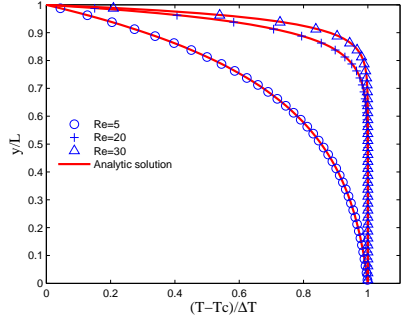}
\caption{Temperature profiles at  $\text{Re}=5, 20$ and $30$ of the porous plate problem with $\text{Pr}=0.71, \text{Ra}=100$. }\label{multit}
\end{figure}
\begin{figure}[!htb]
\centering
\includegraphics[width=0.5\textwidth]{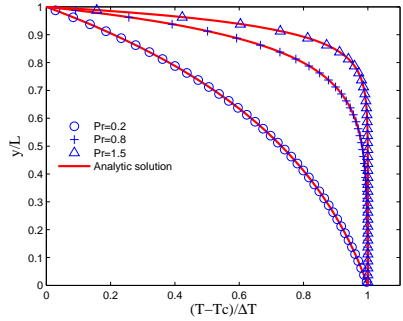}
\caption{Temperature profiles at $\text{Pr}=0.2, 0.8$ and $1.5$ of the porous plate problem with $ \text{Re}=10, \text{Ra}=100 $.}\label{multipr}
\end{figure}

In addition, a set of simulations with variable $\text{Re}$ and $\text{Pr}$ are also carried out to validate the new model.  Uniform mesh with resolution of $N_x\times N_y=80\times40$ is employed, and $RT_0$ is set to be 1.0 so that the flow can be regarded as incompressible. Figs.~\ref{multiv} and \ref{multit} show the normalized velocity and the temperature profiles respectively for $\text{Pr}=0.71$ and $\text{Ra}=100$ with three sets of Reynolds numbers ( $\text{Re}=5, 20$ and $30$). Figures.~\ref{multipr} shows the temperature profiles for $\text{Ra}=100$ and $\text{Re}=10$ with three sets of Prandtl numbers ( $\text{Pr}=0.2, 0.8$ and $1.5$). The analytic solutions are also included for comparison. As shown, the numerical results are in excellent agreement with the analytic ones.
\subsection{Rayleigh-B\'{e}nard Convection}
\begin{figure}[!htb]
\centering
\includegraphics[width=0.5\textwidth]{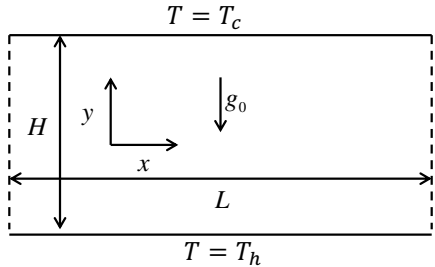}
\caption{Schematic of the Rayleigh-b\'{e}nard convection.}\label{rayleigh}
\end{figure}
The Rayleigh-B\'{e}nard convection is a typical Boussinesq flow. As sketched in Fig.~\ref{rayleigh}, the flow domain is a rectangle with length $L$ and height $H$ $(L=2H)$ and a hot wall on the bottom and a cool wall on the top. The Rayleigh number of this flow is defined by $\text{Ra}=g_0\beta\Delta T H^3/(\nu\kappa)$ where $\Delta T$ is the temperature difference between the hot and the cool walls.

In the simulations, $\Delta T$ is set to be 1.0, and $RT_0$ is set to be 10 so that the code works in the nearly incompressible regime;  boundary conditions developed in Sec.~\ref{boundary} are applied to the top and bottom walls, and periodic boundary conditions are applied to the two side boundaries; the Prandtl number is fixed at $0.71$ (air) in all cases.

\begin{table}[!htb]
\centering
\caption{\label{tablea} Critical Rayleigh numbers with different mesh resolutions. }
\begin{tabular}[t]
   {c c c l l}
   \hline

  & $\text{Mesh}$                &  \hspace{3cm} $\text{Ra}_c$    &\hspace{3cm}  $\text{Error}$ & \\
\hline
   & $40 \times 20$              &  \hspace{3cm}  $1728.68$       &\hspace{3cm}  $1.22\%$  & \\
   & $80 \times 40$              &  \hspace{3cm}  $1711.50$       &\hspace{3cm}  $0.22\%$  &    \\
   & $160 \times 80$             &  \hspace{3cm}  $1706.00$       &\hspace{3cm}  $0.11\%$   &   \\
   & $\text{Theory} $\cite{reid1958some}$$  & \hspace{3cm}  $1707.76$       &  \\

  \hline

\end{tabular}
\end{table}

\begin{figure}[!htb]
\centering
\includegraphics[width=0.5\textwidth]{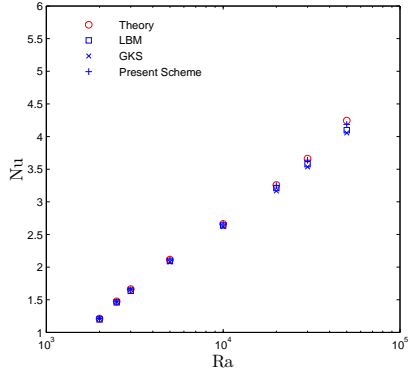}
\caption{The dependence of average Nusselt number on Rayleigh number. The simulation results by the theoretical analysis \cite{clever1974transition}, the original GKS \cite{xu1999rayleigh}, and the LBE method \cite{he1998novel} are also included .}\label{Ra_nu}
\end{figure}

 We first measure the critical Rayleigh number $\text{Ra}_c$, at which the static conductive state becomes unstable.
 Computations are started from the static conductive state at two Rayleigh numbers $1720$ and $1735$ close to $\text{Ra}_c$. An initial small perturbation is applied to the initial temperature field. The growth rates of the disturbance are measured and extrapolated to obtain the critical Rayleigh number corresponding to zero growth rate.  Table \ref{tablea} compiles the calculated critical Rayleigh number on different mesh resolutions and its relative errors compared with the result  obtained by linear stability theory \cite{reid1958some}. It is clearly observed that the coupled DUGKS gives an accurate prediction for the critical Rayleigh number.

Once the Rayleigh-B\'{e}nard convection is stabilized, the heat transfer between the top and bottom is greatly enhanced, which can be quantified by the volume average Nusselt number,
\begin{equation}
\overline{\text{Nu}}=1+\frac{\langle vT \rangle}{\kappa \Delta T/H},
\end{equation}
where $v$ is the vertical velocity and $\langle\cdot\rangle$ represents the average over the whole flow domain.
 Figure.~\ref{Ra_nu} illustrates the calculated relationship between the average Nusselt number and the Rayleigh number with mesh resolution of $80\times40$.  Also included are the results given by theoretical analysis \cite{clever1974transition}, the original GKS \cite{xu1999rayleigh}, and the LBE \cite{he1998novel}. As is shown, our results agree well with theoretical results \cite{clever1974transition} and is slightly better than those of the LBE and GKS methods at high Rayleigh numbers with the same mesh resolution.
 \begin{figure}[!htb]
 \centering
 \begin{subfigure}[]{0.95\linewidth}
 \centering
\includegraphics[width=0.5\textwidth]{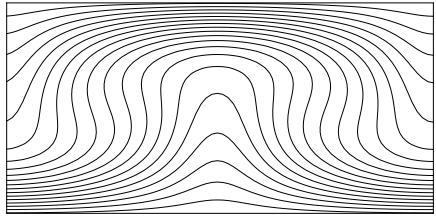}
\caption{Ra=5,000}
\end{subfigure}
\begin{subfigure}[]{0.95\linewidth}
 \centering
\includegraphics[width=0.5\textwidth]{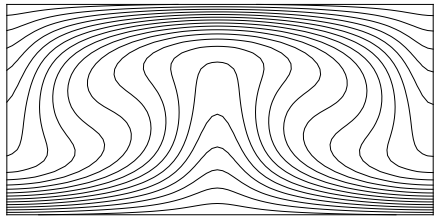}
\caption{Ra=10,000}
\end{subfigure}
\begin{subfigure}[]{0.95\linewidth}
 \centering
\includegraphics[width=0.5\textwidth]{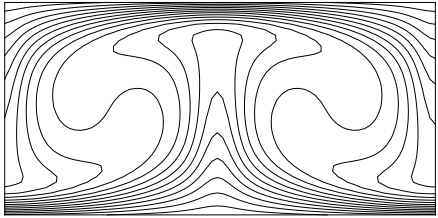}
\caption{Ra=50,000}
\end{subfigure}
\caption{Isothermals of the Rayleigh-B\'{e}nard convection at different Rayleigh numbers.}\label{fig:temperature_Ra}
\end{figure}

\begin{figure}[!htb]
 \centering
 \begin{subfigure}[]{0.95\linewidth}
 \centering
\includegraphics[width=0.5\textwidth]{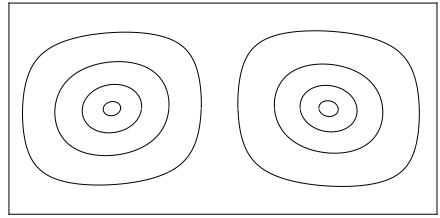}
\caption{\text{Ra}=5,000}
\end{subfigure}
\begin{subfigure}[]{0.95\linewidth}
 \centering
\includegraphics[width=0.5\textwidth]{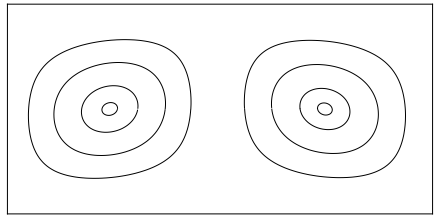}
\caption{\text{Ra}=10,000}
\end{subfigure}
\begin{subfigure}[]{0.95\linewidth}
 \centering
\includegraphics[width=0.5\textwidth]{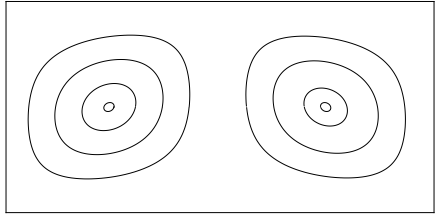}
\caption{\text{Ra}=50,000}
\end{subfigure}
\caption{Streamlines of the Rayleigh-B\'{e}nard convection at different Rayleigh numbers.}\label{fig:velocity_Ra}
\end{figure}

In order to understand the flow characters at different Rayleigh numbers, the isothermals  and the streamlines  at final steady states defined by Eq.~\eqref{work_state} are presented in Figs.~\ref{fig:temperature_Ra} and ~\ref{fig:velocity_Ra}, respectively. It can be seen that as the Rayleigh number increases, two trends occur for the temperature distribution: enhanced mixing of the hot and cold fluids, and an increase in the temperature gradients near the bottom and top boundaries. Both trends enhance the heat transfer in the channel. These phenomena predicted by present model are well consistent with the results in the literatures \cite{shan1997simulation,xu1999rayleigh}.

\subsection{Natural convection in a square cavity}

\begin{figure}
\centering
\includegraphics[width=0.4\textwidth]{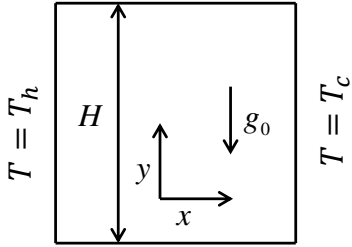}
\caption{Schematic illustration of the flow domain for the natural convection in a square cavity.}\label{square_cavity}
\end{figure}

Natural convection in a square cavity is another canonical test case to validate the thermal model for Boussinesq flows. As illustrated in the Fig.~\ref{square_cavity}, the configuration of the problem considered is a two-dimensional square cavity with a hot wall on the left side and a cool wall on the right. The Rayleigh number of the flow is defined by $\text{Ra}=g_0\beta\Delta T H^3/(\nu\kappa)$ where $\Delta T$ is the temperature difference between the hot and the cool wall, $H$ is the height or width of the cavity.

We first consider the laminar natural convection where the Rayleigh number is less than $10^{6}$. In the simulations, an uniform mesh of $128 \times 128$ points is employed and $RT_0$ is set to be $10$ to make the flow satisfy the incompressible limit. In addition, all the velocities obtained are normalized with the referenced velocity $U_0=\kappa/H$. The velocity boundary condition, Eqs.~\eqref{bounce_flow_F}, is applied to the four walls, the temperature boundary conditions, Eq.~\eqref{bounce_Tc} and \eqref{bounce_Ta}, are applied to the horizontal and vertical walls, respectively.

\begin{figure}[!htb]
\begin{tabular}{cc}
\includegraphics[width=0.4\textwidth]{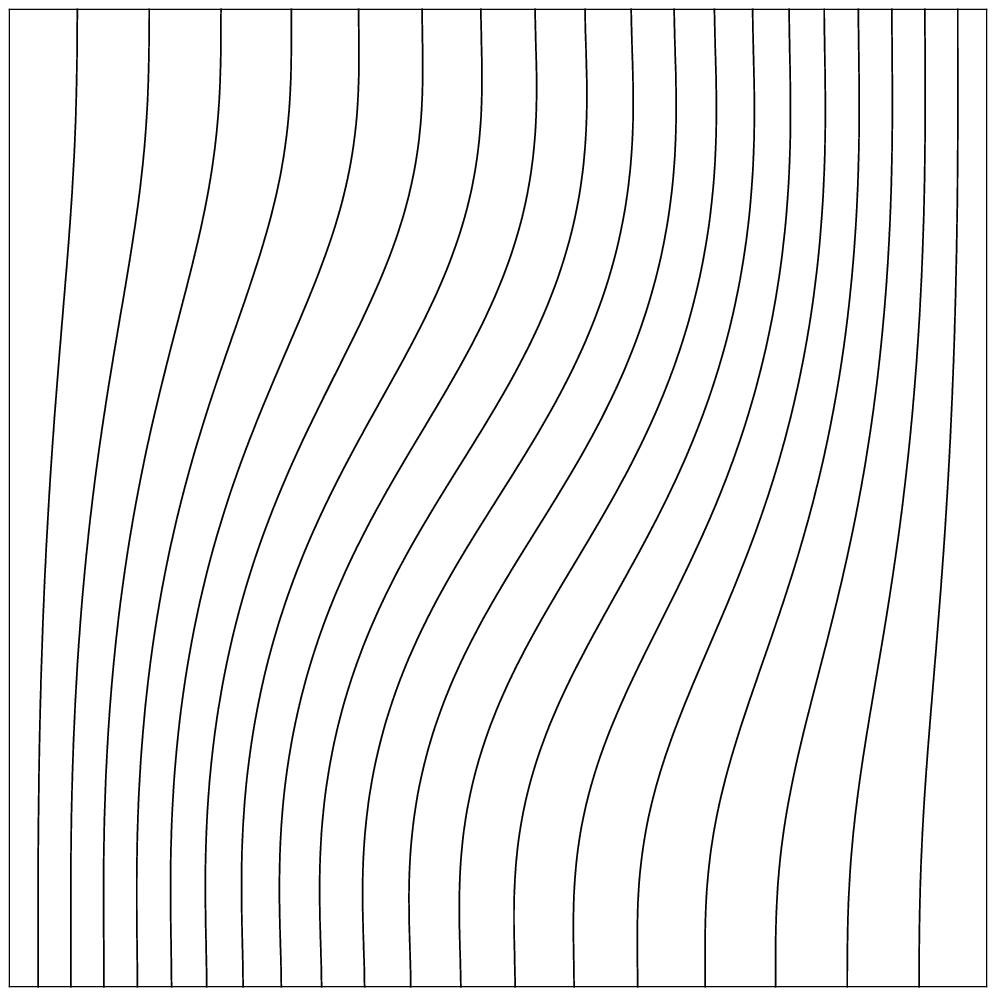}&
\includegraphics[width=0.4\textwidth]{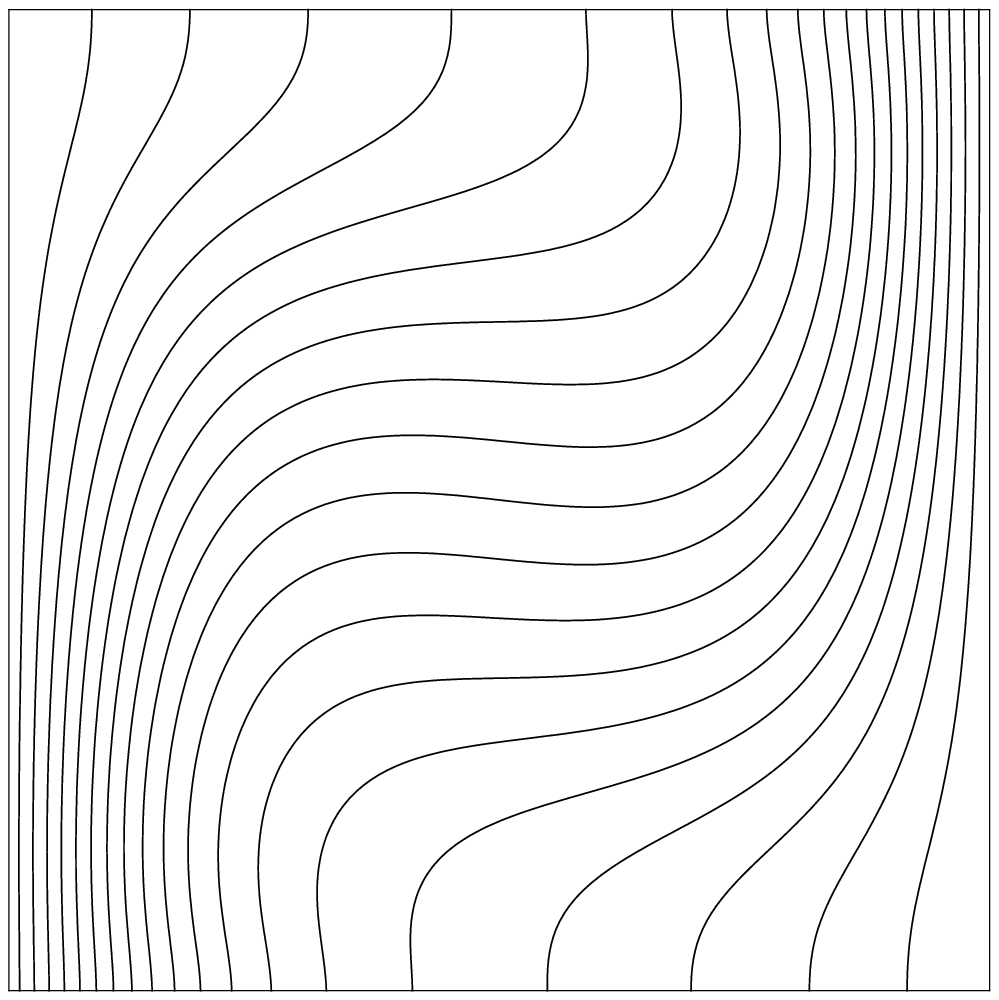}\\
(a)&(b)\\
\includegraphics[width=0.4\textwidth]{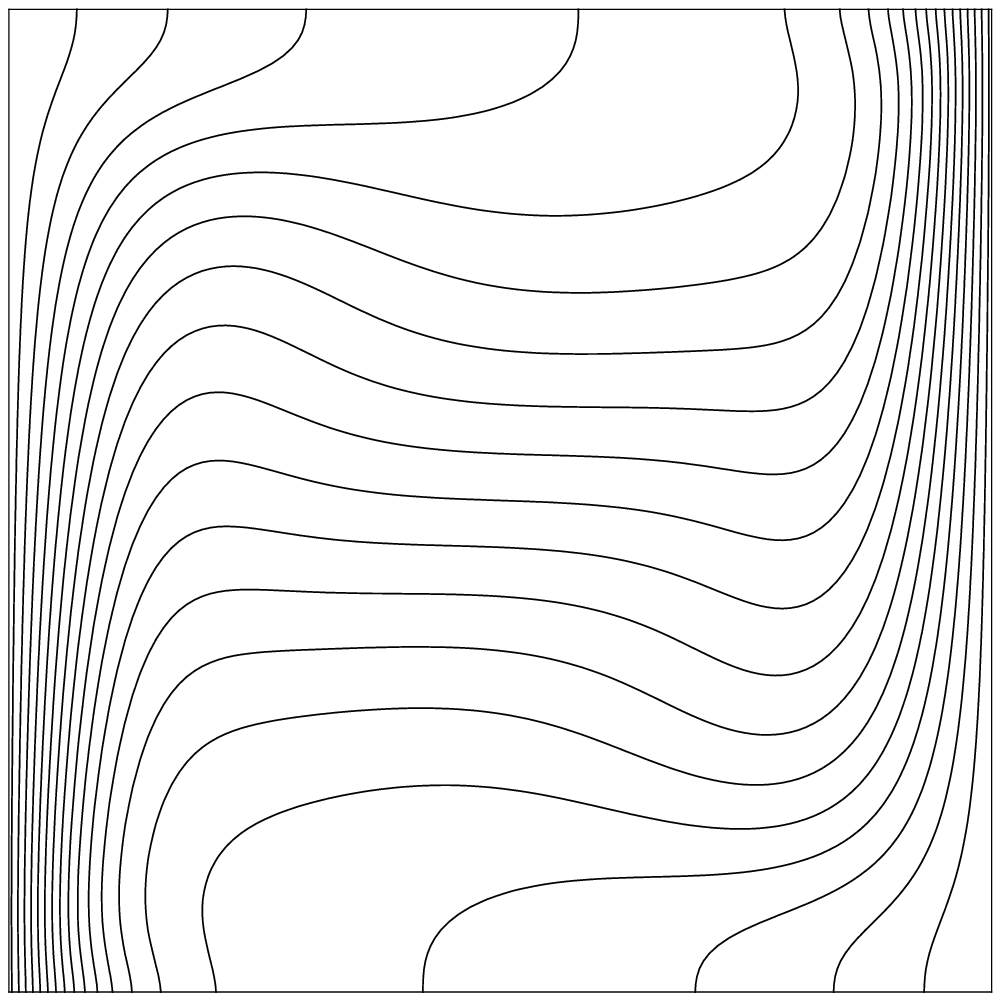}&
\includegraphics[width=0.4\textwidth]{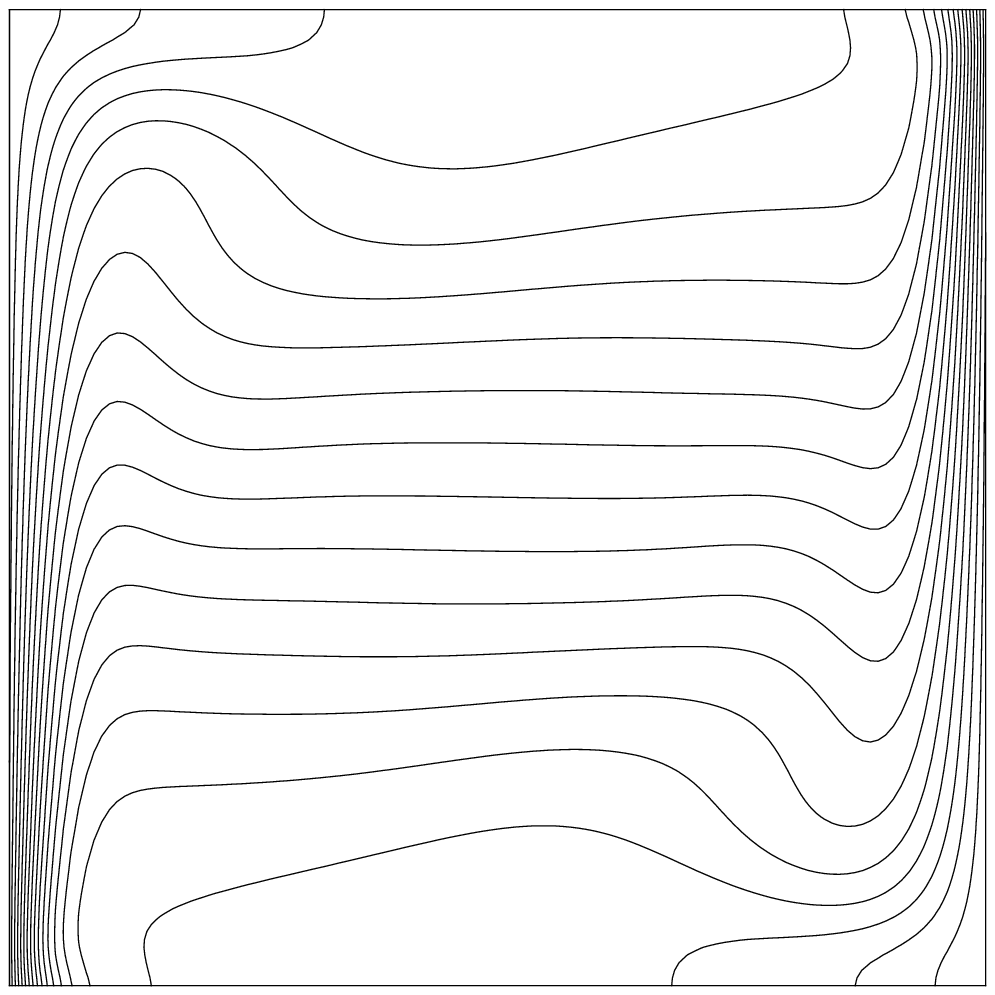}\\
(c)&(d)\\
\end{tabular}
\caption{ Isothermals of the natural convection for (\textit{a}) $\text{Ra}=10^3$,  (\textit{b}) $\text{Ra}=10^4$, (\textit{c}) $\text{Ra}=10^5$ and (\textit{d}) $\text{Ra}=10^6$ with a uniform mesh of $128\times128$ points.}
\label{nc_temperature}
\end{figure}

\begin{figure}[!htb]
\begin{tabular}{cc}
\includegraphics[width=0.4\textwidth]{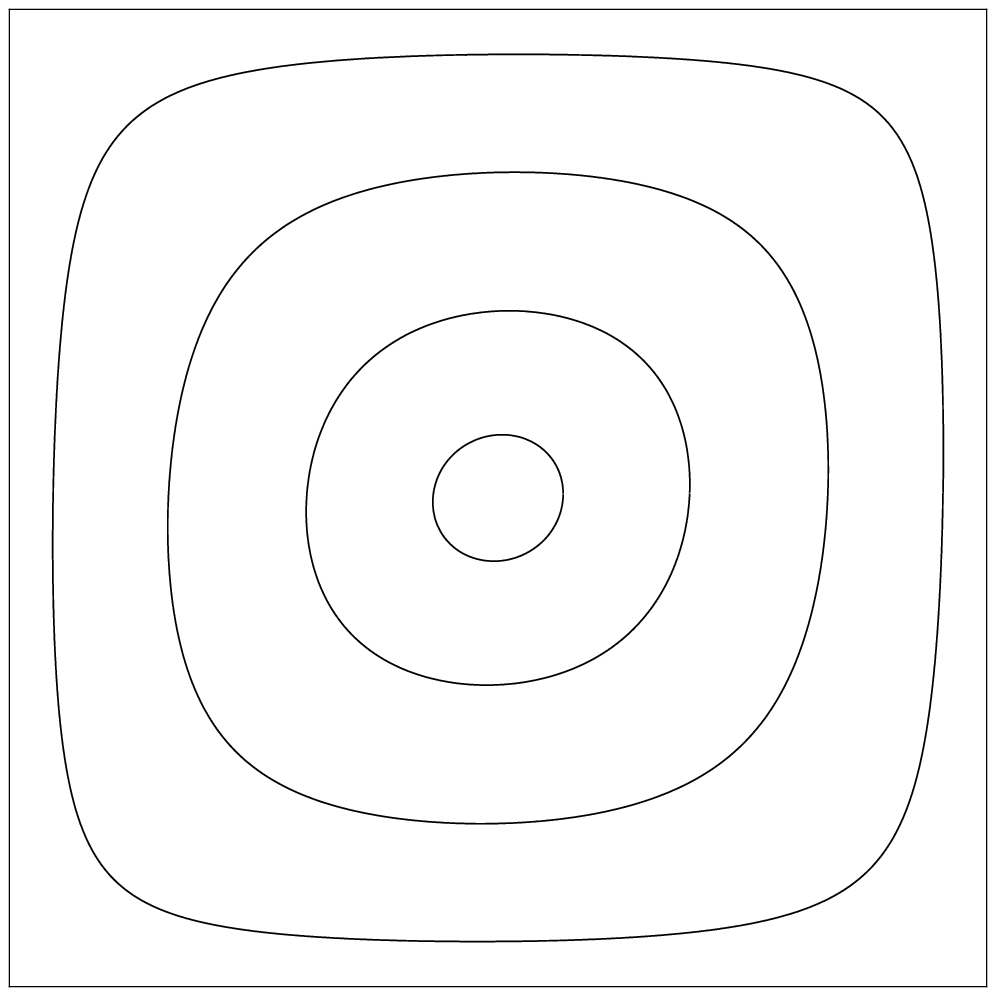}&
\includegraphics[width=0.4\textwidth]{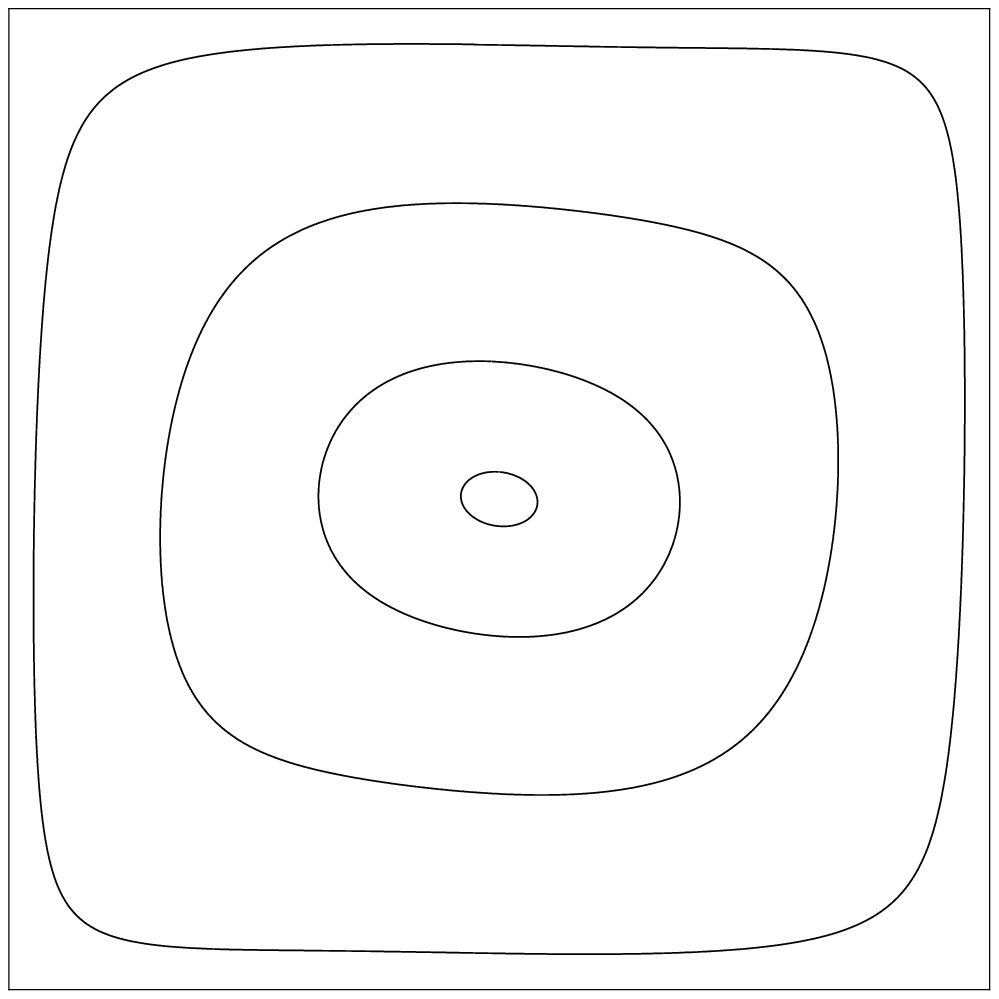}\\
(a)&(b)\\
\includegraphics[width=0.4\textwidth]{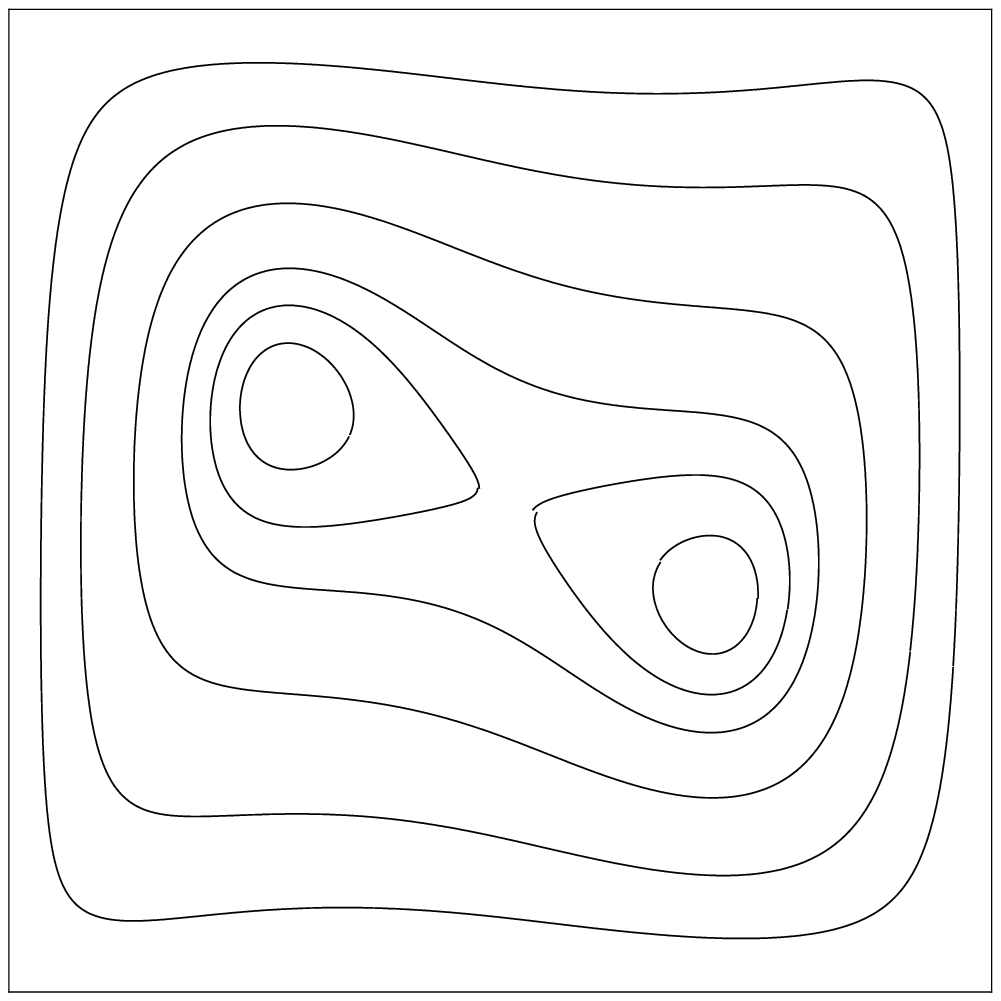}&
\includegraphics[width=0.4\textwidth]{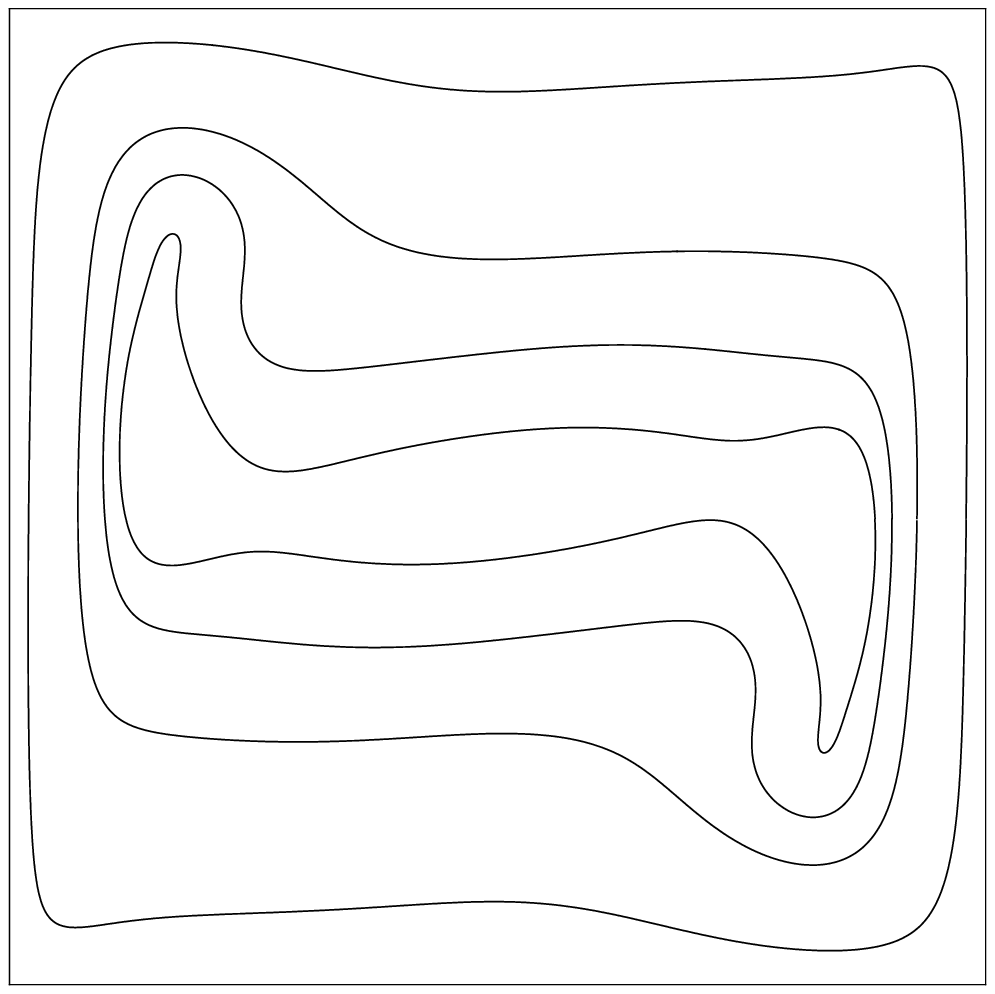}\\
(c)&(d)\\
\end{tabular}
\caption{ Streamlines of the natural convection for (\textit{a}) $\text{Ra}=10^3$,  (\textit{b}) $\text{Ra}=10^4$, (\textit{c}) $\text{Ra}=10^5$ and (\textit{d}) $\text{Ra}=10^6$ with a uniform mesh of $128\times128$ points.}
\label{nc_velocity}
\end{figure}

Figures.~\ref{nc_temperature} and \ref{nc_velocity} show  isothermals and streamlines for $\text{Ra}=10^3, 10^4,
10^5$ and $10^6$ under the stead state defined by Eq.~\eqref{work_state}. It can be seen that as Ra increases, isotherms change from almost vertical to be horizontal in the center of the cavity, and are vertical only in the thin boundary layers near the hot and cold walls. It means that the dominant heat transfer mechanism changes from conduction to convection. Correspondingly, a central vortex appears as the typical features of the flow. The vortex tends to become elliptic and breaks up into two vortices as Ra increases. All those phenomena are agree well with those reported in the literatures \cite{guo2002coupled,wang2013lattice}.

To quantify the results, we compare some quantities of interest with the benchmark results, including the maximum horizontal velocity on the vertical centerline of the cavity, $u_{max}$, and the corresponding $y-$coordinate,  the maximum vertical velocity  on the horizontal centerline of the cavity, $v_{max}$, and the corresponding $x-$coordinate, the maximum value of local Nusselt number on the cool  boundary, $\text{Nu}_{max}$, and the corresponding $y-$coordinate, and the averaged Nusselt number throughout the cavity $\overline{\text{Nu}}$. Table \ref{tablec} compares the predictions from the present calculations with the literature results, also included are the relative errors. It is clearly observed that the present results are in excellent agreement with the reference solutions.

\begin{table}[!htb]
\caption{\label{tablec} Comparison of the numerical results of the coupled DUGKS with those reported in the literature \cite{de1983natural}.}
\centering
\begin{tabular}[t]
   {l l l l l l l l l}
   \hline

   &$\text{Ra}$  &     \hspace{4mm}$         $ &\hspace{4mm} $10^3  $    &\hspace{4mm}  $10^4   $  &\hspace{4mm} $10^5$   & \hspace{4mm} $10^6$&    \\
  \hline
   &$u_{max}$    &     \hspace{4mm} $\text{Present}$ &\hspace{4mm} $3.6445$    &\hspace{4mm}  $16.1737$ &\hspace{4mm}  $34.8016$  & \hspace{4mm}  $63.758$  &\\
   &$       $    &     \hspace{4mm} $$\cite{de1983natural}$$ &\hspace{4mm} $3.649 $    &\hspace{4mm}  $16.178 $ &\hspace{4mm}  $34.73$   & \hspace{4mm}  $64.63$   &\\
   &$\text{Error}(\%)$&\hspace{4mm} $        $ &\hspace{4mm} $0.123 $    &\hspace{4mm}  $0.025  $ &\hspace{4mm}  $0.223$  & \hspace{4mm}  $1.35$    &\\
   &$  y    $    & \hspace{4mm}     $\text{Present}$ &\hspace{4mm} $0.8203$    &\hspace{4mm}  $0.8281 $ &\hspace{4mm}  $0.8594$  & \hspace{4mm}  $0.8594$  &\\
   &$       $    & \hspace{4mm}     $$\cite{de1983natural}$$ &\hspace{4mm} $0.813 $    &\hspace{4mm}  $0.823  $ &\hspace{4mm}  $0.855$  & \hspace{4mm}  $0.850$   &\\
   &$\text{Error}(\%)$&\hspace{4mm} $        $ &\hspace{4mm} $0.898 $    &\hspace{4mm}  $0.620  $ &\hspace{4mm}  $0.515$  & \hspace{4mm}  $1.11$    &\\
   &$v_{max}$    &\hspace{4mm}      $\text{Present}$ &\hspace{4mm} $3.6989$    &\hspace{4mm}  $19.6194$ &\hspace{4mm}  $68.3318$  & \hspace{4mm}  $215.5081$&\\
   &$       $    &\hspace{4mm}      $$\cite{de1983natural}$$ &\hspace{4mm} $3.697 $    &\hspace{4mm}  $19.617 $ &\hspace{4mm}  $68.59$   & \hspace{4mm}  $219.36$  &\\
   &$\text{Error}(\%)$&\hspace{4mm} $        $ &\hspace{4mm} $0.0514$    &\hspace{4mm}  $0.012  $ &\hspace{4mm}  $0.376$  & \hspace{4mm}  $1.76$    &\\
   &$  x    $    &\hspace{4mm}      $\text{Present}$ &\hspace{4mm} $0.1797$    &\hspace{4mm}  $0.125  $ &\hspace{4mm}  $0.0703$ & \hspace{4mm}  $0.0391$  &\\
   &$       $    &\hspace{4mm}      $$\cite{de1983natural}$$ &\hspace{4mm} $0.178 $    &\hspace{4mm}  $0.119  $ &\hspace{4mm}  $0.066$  & \hspace{4mm}  $0.0379$  &\\
   &$\text{Error}(\%)$&\hspace{4mm} $        $ &\hspace{4mm} $0.952 $    &\hspace{4mm}  $5.04   $ &\hspace{4mm}  $6.515$  & \hspace{4mm}  $3.17$    &\\
   &$\text{Nu}_{max}$ &\hspace{4mm} $\text{Present}$ &\hspace{4mm} $1.5080$    &\hspace{4mm}  $3.5313 $ &\hspace{4mm}  $7.6555$  & \hspace{4mm}  $16.6194$&\\
   &$       $    &\hspace{4mm}      $$\cite{de1983natural}$$ &\hspace{4mm} $1.505 $    &\hspace{4mm}  $3.328  $ &\hspace{4mm}  $7.717$   & \hspace{4mm}  $17.925$  &\\
   &$\text{Error}(\%)$&\hspace{4mm} $        $ &\hspace{4mm} $0.199 $    &\hspace{4mm}  $0.094  $ &\hspace{4mm}  $0.796$  & \hspace{4mm}  $7.28$    &\\
   &$  y    $    &\hspace{4mm}      $\text{Present}$ &\hspace{4mm} $0.9141$    &\hspace{4mm} $0.8594  $ &\hspace{4mm}  $0.9219$ & \hspace{4mm}  $0.9531$  &\\
   &$       $    &\hspace{4mm}      $$\cite{de1983natural}$$ &\hspace{4mm} $0.908 $    &\hspace{4mm} $0.857   $ &\hspace{4mm}  $0.919$  & \hspace{4mm}  $0.9622$  &\\
   &$\text{Error}(\%)$&\hspace{4mm} $        $ &\hspace{4mm} $6.72  $    &\hspace{4mm}  $0.280  $ &\hspace{4mm}  $0.316$  & \hspace{4mm}  $0.946$    &\\
   &$\overline{\text{Nu}}$&\hspace{4mm} $\text{Present}$ & \hspace{4mm} $1.1181$&\hspace{4mm}  $2.2431$ &\hspace{4mm}  $4.5024$  & \hspace{4mm}  $8.6796$  &\\
   &$       $    &\hspace{4mm}      $$\cite{de1983natural}$$ &\hspace{4mm} $1.118 $    &\hspace{4mm}  $2.243  $ &\hspace{4mm}  $4.519$  & \hspace{4mm}  $8.8$     &\\
   &$\text{Error}(\%)$&\hspace{4mm} $        $ &\hspace{4mm} $0.00894$   &\hspace{4mm}  $0.004  $ &\hspace{4mm}  $0.367$  & \hspace{4mm}  $1.37$    &\\

  \hline

\end{tabular}
\end{table}

\begin{figure}[!htb]
\centering
\includegraphics[width=0.49\textwidth,height=0.3\textheight]{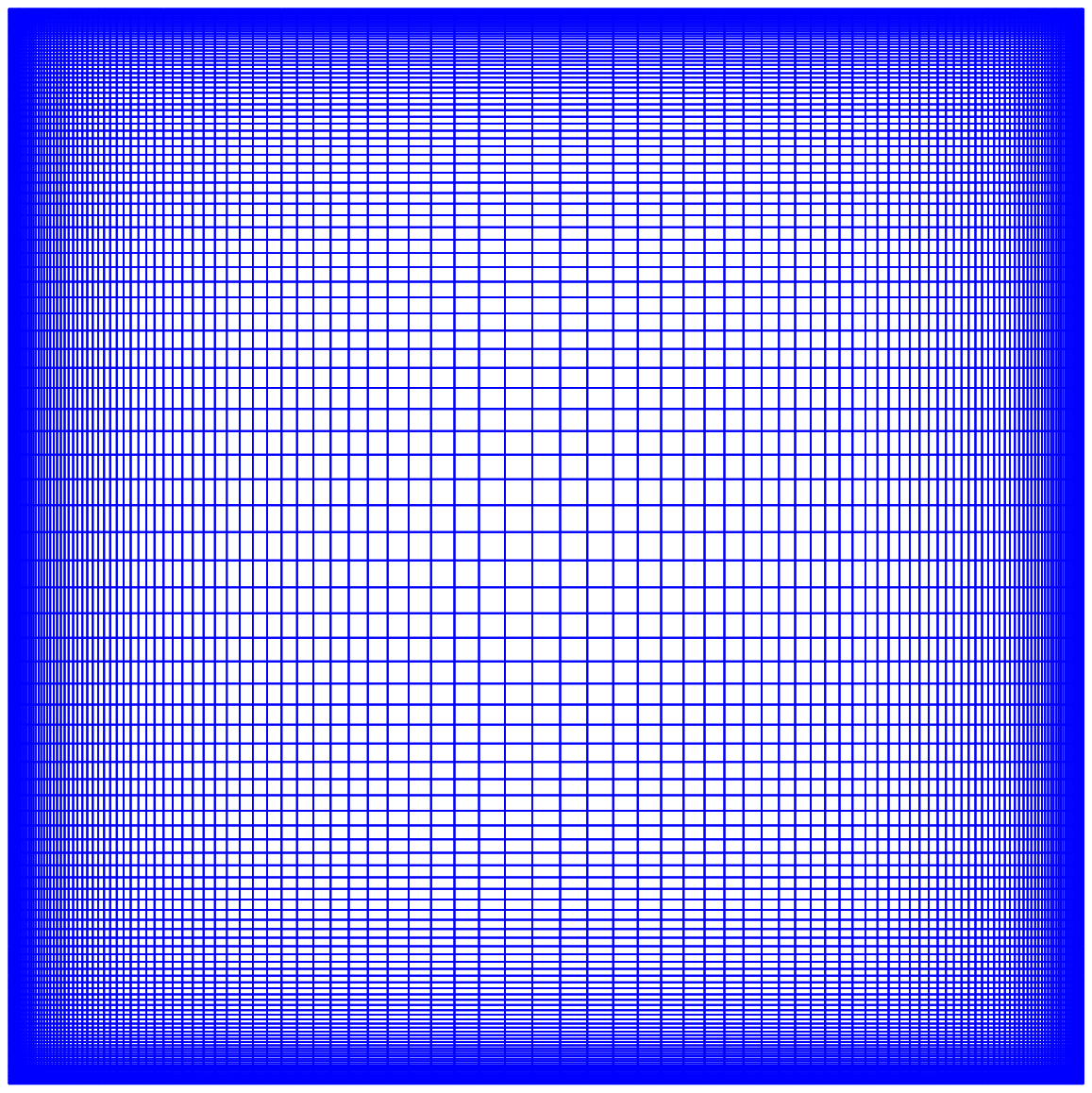}
\caption{Numerical mesh with $128\times128$ grid points for the turbulent natural convection calculations }\label{mesh}
\end{figure}

In order to further test the capability of the present DUGKS for simulating thermal flows, we now apply it to simulate the turbulent natural convection at high Rayleigh numbers. In such cases, thinning boundary layers along the hot and cool walls occur in which steeper velocity and temperature gradients appear, and fine mesh resolutions should be used. The FV nature of the DUGKS makes it easy to vary the mesh resolution according to the local accuracy requirement. In the current test, a non-uniform mesh is adopted, as shown in Fig.~\ref{mesh}, where the mesh resolution follows a geometric progression for the grid spacing.

\begin{figure}[!htb]
\begin{tabular}{cc}
\includegraphics[width=0.4\textwidth]{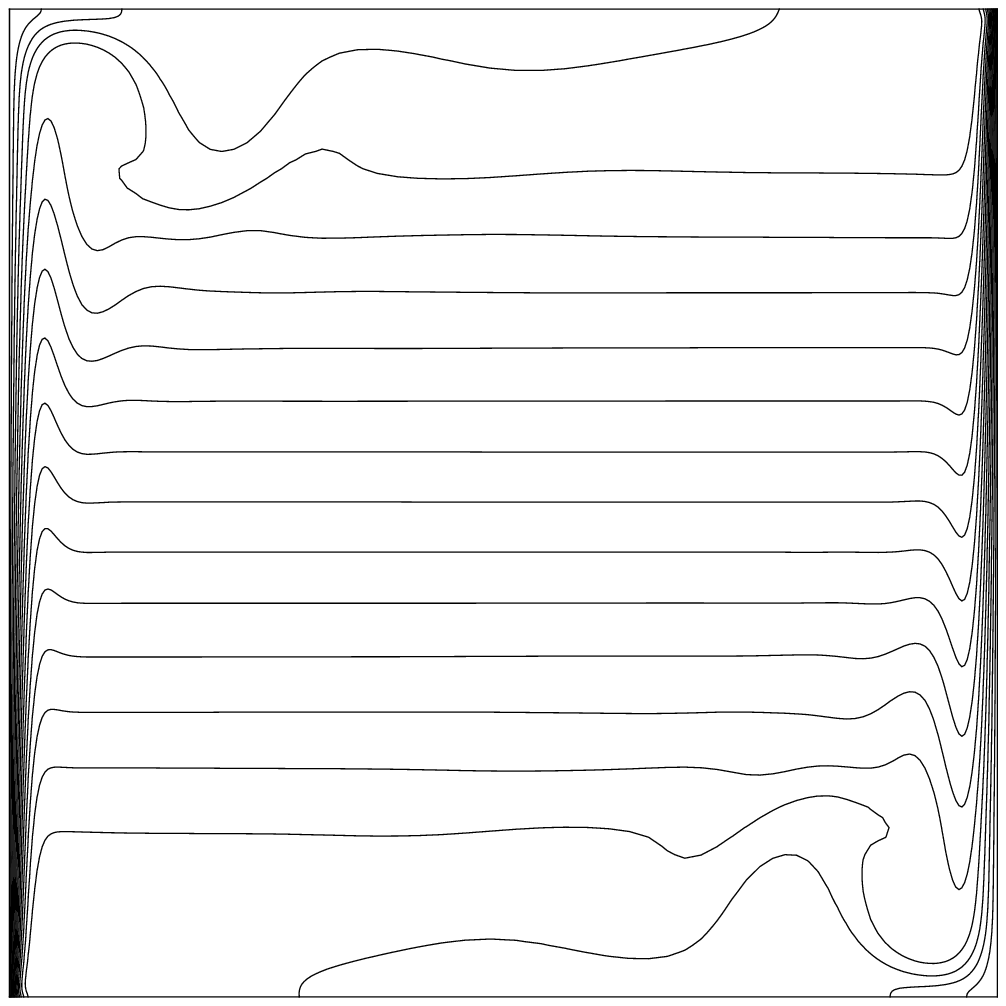}&
\includegraphics[width=0.4\textwidth]{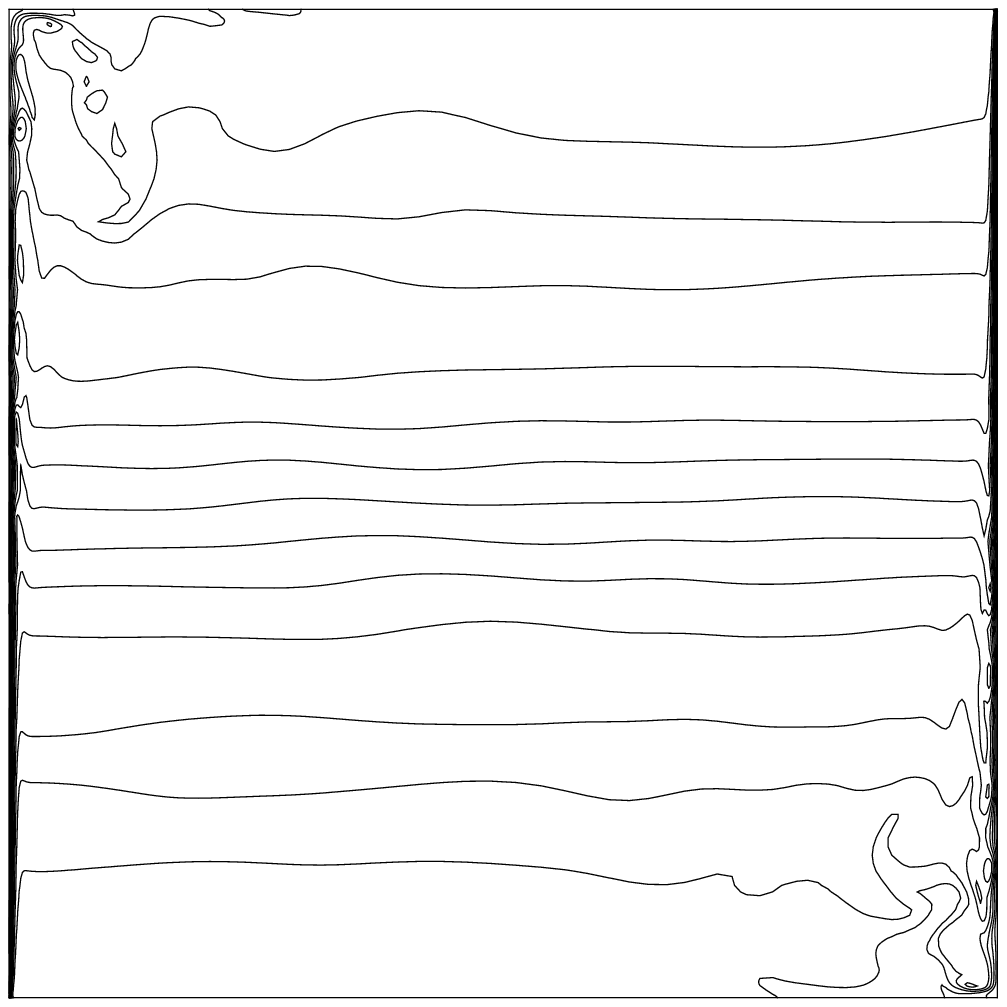}\\
(a)&(b)\\
\includegraphics[width=0.4\textwidth]{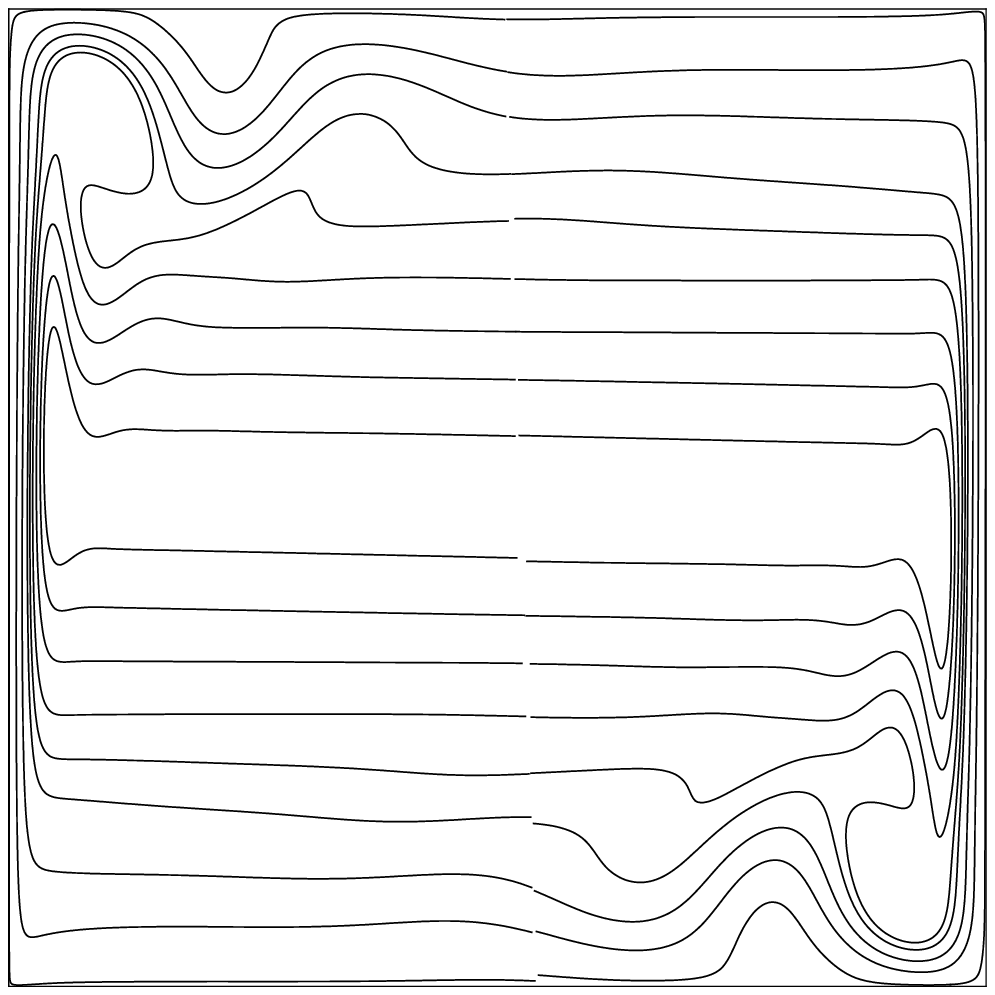}&
\includegraphics[width=0.4\textwidth]{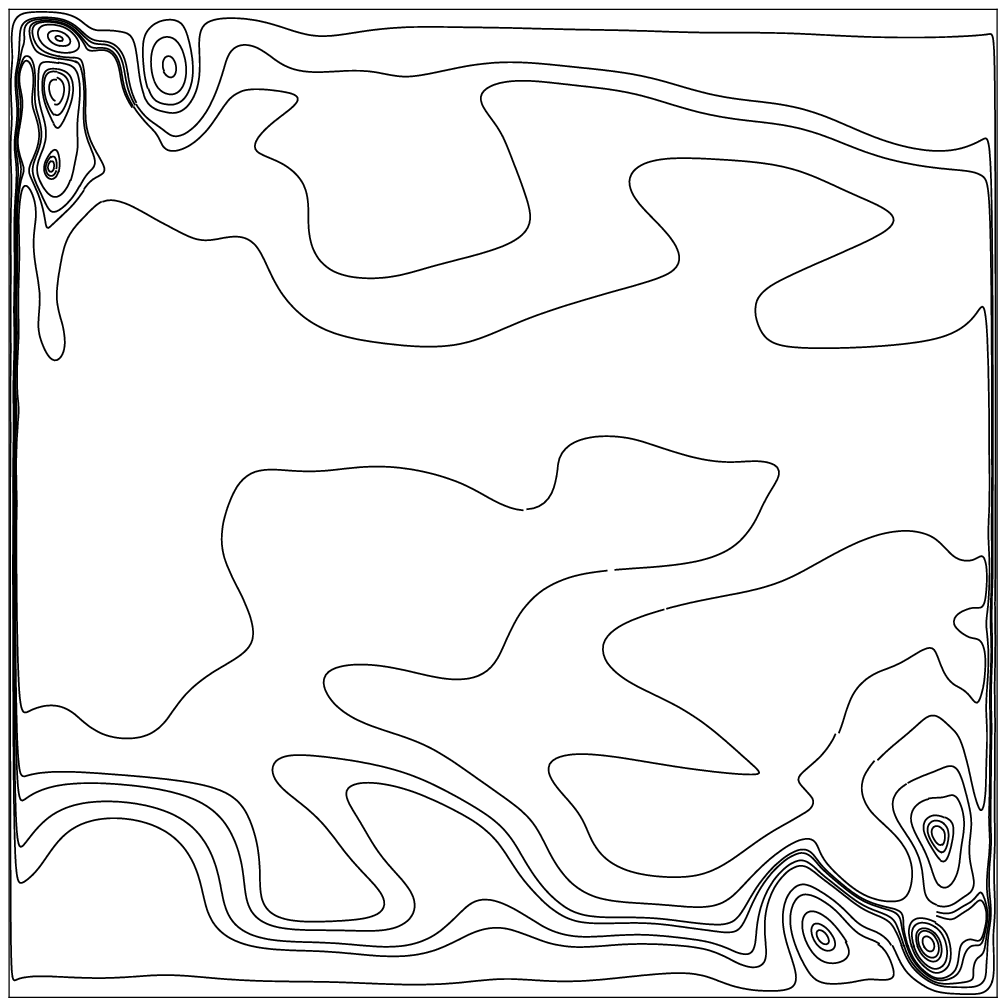}\\
(c)&(d)\\
\end{tabular}
\caption{  Isothermals for (\textit{a}) $\text{Ra}=10^8$, and (\textit{b}) $\text{Ra}=10^{10}$, and streamlines  for (\textit{c}) $\text{Ra}=10^8$, and (\textit{d}) $\text{Ra}=10^{10}$ in the natural convection with a non-uniform mesh of $128\times128$ points.}
\label{tur_velocity}
\end{figure}

Figure.~\ref{tur_velocity} shows instantaneous isothermals and streamlines at $\text{Ra}=10^8$ and $10^{10}$.
As show, at $\text{Ra}=10^8$, the isothermals are horizontal at the center of the cavity and become vertical near the hot and cool walls; the vortices appear at the top-left and lower-right corner due to the fast moving of the fluid near the walls. As Ra increase to $10^{10}$, the isotherms at the center region of cavity are not  straight longer and present a wavy state, while become irregular at the top-left and bottom-right corners of cavity; and the small-scale vortices occur in  entire simulation domain and the flow structure becomes irregular and chaotic.
All those observations agree well with the previous results \cite{dixit2006simulation,zhuo2013based,contrino2014lattice}.

Quantificationally, we measure some time-average qualities \cite{zhuo2013based}, such as the maximum horizontal velocity on the vertical centerline of the cavity, $u_{max}$, and the corresponding $y-$coordinate, the maximum vertical velocity on the horizontal centerline of the cavity, $v_{max}$, and the corresponding $x-$coordinate, and the average Nusselt number on the hot wall ($\overline{\text{Nu}})$ . Table \ref{tablee} gives the results calculated by the present method, also included are the results obtained from the pseudo-spectral method \cite{le1991accurate} and the LBE method with finer mesh resolutions \cite{dixit2006simulation}. As shown in the table, the obtained results are  in good agreement with those reported in the literatures.
It must be emphasized that no turbulence model is employed in the present simulation and the FV nature of DUGKS makes it easy to adopt non-uniform meshes.

\begin{table}[!htb]
\centering
\caption{\label{tablee} Comparison of the numerical results of the present work with those reported in the literatures.}
\begin{tabular}[t]
   {c c l l l l l l}

   \hline

   & $\text{Ra}          $  &  \hspace{2cm}  $          $    &  \hspace{2cm}    $10^8$           &\hspace{2cm} $10^{10}$ &\\
   \hline
   & $\overline{u}_{max} $  &  \hspace{2cm}  $\text{Present}  $    &  \hspace{2cm}    $332.2619  $     &\hspace{2cm}  $2334.7$  & \\
   & $                   $  &  \hspace{2cm}  $$\cite{le1991accurate}$$    &  \hspace{2cm}    $321.876   $     &\hspace{2cm}  $-$ & \\
   & $                   $  &  \hspace{2cm}  $$\cite{dixit2006simulation}$$    &  \hspace{2cm}    $389.877   $     &\hspace{2cm}  $2323$ & \\
   & $y_{max}            $  &  \hspace{2cm}  $\text{Present}  $    &  \hspace{2cm}    $0.9396    $     &\hspace{2cm}  $0.9436$  & \\
   & $                   $  &  \hspace{2cm}  $$\cite{le1991accurate}$$    &  \hspace{2cm}    $0.928     $     &\hspace{2cm}  $-$ & \\
   & $                   $  &  \hspace{2cm}  $$\cite{dixit2006simulation}$$    &  \hspace{2cm}    $0.937     $     &\hspace{2cm}  $0.9423$ & \\
   & $\overline{v}_{max} $  &  \hspace{2cm}  $\text{Present}  $    &  \hspace{2cm}    $2229.7    $     &\hspace{2cm}  $22282$  & \\
   & $                   $  &  \hspace{2cm}  $$\cite{le1991accurate}$$    &  \hspace{2cm}    $2222.39   $     &\hspace{2cm}  $-$ & \\
   & $                   $  &  \hspace{2cm}  $$\cite{dixit2006simulation}$$    &  \hspace{2cm}    $2231.374  $     &\hspace{2cm}  $21463$ & \\
   & ${x}_{max}          $  &  \hspace{2cm}  $\text{Present}  $    &  \hspace{2cm}    $0.0123    $     &\hspace{2cm}  $0.0039$  & \\
   & $                   $  &  \hspace{2cm}  $$\cite{le1991accurate}$$    &  \hspace{2cm}    $0.012     $     &\hspace{2cm}  $-$ & \\
   & $                   $  &  \hspace{2cm}  $$\cite{dixit2006simulation}$$    &  \hspace{2cm}    $0.0112    $     &\hspace{2cm}  $0.0049$ & \\
   & $\overline{Nu}      $  &  \hspace{2cm}  $\text{Present}  $    &  \hspace{2cm}    $30.5041   $     &\hspace{2cm}  $103.7467$  & \\
   & $                   $  &  \hspace{2cm}  $$\cite{le1991accurate}$$    &  \hspace{2cm}    $30.225    $     &\hspace{2cm}  $-$ & \\
   & $                   $  &  \hspace{2cm}  $$\cite{dixit2006simulation}$$    &  \hspace{2cm}    $30.506    $     &\hspace{2cm}  $103.663$ & \\

  \hline
\end{tabular}
\end{table}

Numerical instability has been a primary concern in previous thermal kinetic methods. Now, we compare the stability of present model with the well-accepted thermal LBE model (CLBGK) \cite{guo2002coupled} under the same initial state and boundary conditions without considering the accuracy of the results. First, we measure the minimum required mesh resolution at a fixed Rayleigh number under steady state criterion of Eq.~\eqref{work_state}. Table \ref{tabled} shows the minimum required mesh resolution at the given Rayleigh numbers. It can be seen that the DUGKS requires much less mesh points than the CLBGK in order to get a stable solution. For example, even at $\text{Ra}=10^6$, the DUGKS can reach a steady state solution with $10\times10$ uniform mesh points. Second, we evaluate the maximum Rayleigh number on a specific mesh resolution at which the computation is still stable. It is found that with a fixed mesh resolution, the DUGKS can reach a much higher $\text{Ra}$ than the CLBGK. For instance, on a uniform $32\times32$ mesh points, the computation from the CLBGK blows up at $\text{Ra}=10^5$. However, the coupled DUGKS works
even at $\text{Ra}=10^{12}$. Clearly, in comparison with the CLBGK, the DUGKS has super performance in stability. However, in terms of the computational efficiency, the CLBGK is about three times faster than the DUGKS for each node updating per time step due to the additional equilibrium distribution evaluation in DUGKS in order to include the collision effect into its flux transport. This is consistent with previous results for isothermal flows  \cite{wangcomparative}.
\begin{table}[htbp]
\centering
\caption{\label{tabled} The minimum required mesh resolution at different Rayleigh numbers.}
\begin{tabular}[t]
   {c c c c c c c c  c}

   \hline

   & $\text{Ra}  $  &  \hspace{3cm} $\text{CLBGK}$  &\hspace{3cm} $\text{DUGKS}$ &\\
   \hline
   & $10^4$  &  \hspace{3cm} $20\times20$   &\hspace{3cm}  $5\times5$  & \\
   & $10^5$  &  \hspace{3cm} $30\times30$   &\hspace{3cm}  $10\times10$ & \\
   & $10^6$  &  \hspace{3cm} $60\times60$   &\hspace{3cm}  $10\times10$ & \\
  \hline

\end{tabular}
\end{table}

\section{Conclusions}
\label{conclusion}
In this paper,  a coupled discrete unified gas-kinetic scheme is developed for the Boussinesq flows. The velocity field and temperature field are separately described by two distributions, and the DUGKS with an external force term is presented in the DUGKS algorithm.  The simulation results demonstrate that the coupled DUGKS is of second order accuracy, and can accurately describe the laminar and turbulent thermal convection. Particularly, in comparison with the LBE methods, the coupled DUGKS can adopt the non-uniform mesh in a natural way, and has a remarkable performance in terms of the numerical stability .

\section*{ACKNOWLEDGMENT}
 This study is financially supported
by the National Natural Science Foundation of China
(Grant No. 51125024) and the Fundamental Research Funds for the Central
Universities (Grant No. 2014TS119).
\section*{References}
\bibliography{mybibfilea}
\end{document}